\documentclass[aps, prd, preprintnumbers, floatfix, showpacs, superscriptaddress, 10pt]{revtex4}

\usepackage[dvips]{graphics}
\usepackage{amsmath,amsfonts,bm}
\usepackage{epsfig,bm}
\usepackage{graphicx,comment}
\usepackage{dcolumn,color}

\def\cA{{\cal A}}   
   
   \def\cL{{\cal L}}
\def\cM{{\cal M}}  \def\cO{{\cal O}} \def\cP{{\cal P}}
 \def\cR{{\cal R}}  \def\cT{{\cal T}}
 \def\cV{{\cal V}}  
 
\def\tr{\mathop{\rm tr}}  \def\Tr{\mathop{\rm Tr}}  
     
   \def\GeV{{\rm GeV}}       \def\MeV{{\rm MeV}}
                
\def\sla{\slash{\!\!\!}}

\newcommand{\be}{\begin{equation}}
\newcommand{\ee}{\end{equation}}
\newcommand{\bear}{\begin{eqnarray}}
\newcommand{\eear}{\end{eqnarray}}
\newcommand{\ba}{\begin{array}}
\newcommand{\ea}{\end{array}}

\newcommand{\dt}{\partial}

\newcommand\s{\sigma}
\newcommand\eps{\epsilon}
\newcommand\gam{\gamma}
\newcommand\om{\omega}

\begin{document}
\preprint{EFI 09-11}


\title{Photoproduction through Chern-Simons Term Induced Interactions\\ in Holographic QCD}


\author{Sophia~K.~Domokos}
\affiliation{Enrico Fermi Institute and Department of Physics,
University of Chicago, Chicago Illinois 60637, USA}
\author{Hovhannes~R.~Grigoryan}
\affiliation{Physics Division, Argonne National Laboratory, \\
Argonne, IL 60439-4843 USA}
\author{Jeffrey~A.~Harvey}
\affiliation{Enrico Fermi Institute and Department of Physics,
University of Chicago, Chicago Illinois 60637, USA}


\begin{abstract}
We employ both top-down and bottom-up  holographic dual models of QCD to calculate vertex functions and couplings
that are induced by the five dimensional Chern-Simons term.
We use these couplings to study the photoproduction of $f_1$ mesons.  The Chern-Simons-term-induced interaction leads to a simple relation between the polarization of the incoming photon and the final state $f_1$ meson which should
allow a clear separation of this interaction from competing processes.
\end{abstract}



\maketitle


\section{Introduction}

There are many reasons to believe that QCD has a dual description in terms of string theory. Prominent among these, on the theoretical side, are the properties of QCD at large $N_c$ which include an infinite tower of mesons of arbitrarily high spin and a topological expansion which mimics that of string theory. On the experimental side, the observed Regge behavior of strong interaction processes at large center of mass energy $s$ and fixed momentum transfer $t$ is suggestive of string theory, and fits to total cross sections suggest a string description involving both Reggeons (open strings) and a Pomeron (closed strings) \cite{dl}.  There is persuasive new evidence for this duality, as well as a concrete prescription of how it should be implemented in the form of the string/gauge theory correspondence \cite{Maldacena:1997re,Gubser,Witten}.

Although the full dual theory of QCD is not known, models which capture some of the central features of QCD have been developed. These theories include top-down models based on D-brane constructions  \cite{Sakai:2004cn,Sakai:2005yt}, and bottom-up phenomenological models which incorporate the essential features of low-energy QCD into a dual description involving gauge theory in a five-dimensional space \cite{Erlich:2005qh, DaRold:2005zs} (see also Refs.~\cite{Polchinski:2002jw} and \cite{Brodsky:2003px}). These models also have serious flaws \cite{Csaki:2008dt}. Their development into more fully realistic dual theories requires testing their validity in a variety of settings.

Of particular interest to us in this article are the Chern-Simons terms arising in the Lagrangians of dual theories, which are determined by the anomaly structure of QCD. These give rise to pseudo-Chern-Simons couplings between the vector and axial-vector mesons of QCD, as well as more standard couplings appearing in gauged WZW terms. Some consequences of such couplings were explored in \cite{Domokos:2007kt,Harada}. Similar couplings also occur in the Standard Model, where the electroweak gauge bosons have pseudo-Chern-Simons couplings to QCD vector and axial-vector mesons when the WZW
term is gauged in a way that includes the full gauge group and anomaly structure of the Standard Model \cite{Harvey:2007rd, Harvey:2007ca}. Here we develop the calculations needed to compare the consequences of these couplings to the experimental data on photoproduction of vector and axial-vector mesons. We note that although couplings of similar form
can be written down on purely phenomenological grounds, the precise forms we get here are dictated by the general
principles of AdS/QCD and in particular by the demand that the anomalous variation of the dual theory correctly match the flavor anomalies of QCD.

The high intensity electron beam at Thomas Jefferson National Laboratory (JLab) and its large acceptance detector at Hall D make possible a systematic study of vector and axial-vector meson photoproduction at intermediate photon energy $ E_{\gamma} \sim 10 \ \GeV $ and small momentum transfer (forward angles). We focus on the region of small momentum transfer, where perturbative QCD is of little use and the exchange mechanism is dominated by the Regge trajectories in the $t$-channel.

We first consider single particle exchange, using the Chern-Simons term in holographic QCD to compute the anomalous couplings and vertex functions.  Keeping the single-particle vertex structure intact, we then replace the usual  single-particle Feynman propagator by the ``Reggeized propagator,'' which includes contributions from the entire Regge trajectory of the exchanged meson. Especially relevant to JLab data is (axial)vector-meson photoproduction from (hydrogen) nuclei, and we focus on this process here. Though the vertices involving nucleons could also be studied in a holographic framework (see e.g. \cite{Hata:2007mb,Hashimoto:2008zw,Hong:2007ay}), this lies outside the scope of the present work, and we instead take these vertices from other phenomenological models.


The outline of this paper is as follows. In section 2 we introduce the two dual models of QCD which we will use to perform explicit calculations. The first is the bottom-up model introduced in \cite{Hirn:2005nr} following closely related models \cite{Erlich:2005qh,DaRold:2005zs}. The second is the top-down model of Sakai and Sugimoto \cite{Sakai:2004cn}. In section 3 we discuss the anomalous couplings in these models
and use them to derive a set of four-dimensional (4D) couplings for pions, vector mesons and axial-vector mesons. In section 4 we turn to calculations of photoproduction ($\gamma + N \rightarrow {\rm Meson} + N$) with a focus on photoproduction of the $f_1$ meson. This process provides a fairly clean test of the anomalous couplings, which predict a simple correlation between the polarization of the outgoing $f_1$ and the polarization of the incident photon. We first derive single particle exchange diagrams, then discuss their extension to exchange of full Regge trajectories in order to describe the scattering in the Regge regime. At the end of section 4 we outline other processes that may prove relevant in probing the structure of the $f_1$ Regge trajectory. In the final section we conclude and suggest extensions of the current work.
In the three appendices we give a more detailed list of various pseudo-Chern-Simons couplings (Appendix A), a brief analysis of the contribution of nucleon exchange to $f_1$ photoproduction (Appendix B), and a comparison of  the decay $f_1 \rightarrow \rho^0 + \gamma$ using our results to the experimentally measured rate and polarization structure.


\section{Dual Models of QCD}

In this section we review the general features of the AdS/QCD construction and the structure of two proposed QCD duals. This will serve mainly to fix our notation and conventions; additional details may be found in \cite{Hirn:2005nr} and \cite{Sakai:2004cn}.

\subsection{General Structure of AdS/QCD Models}

AdS/QCD models tweak the original AdS/CFT duality between ${\cal N}=4~SYM$ and string theory on $AdS_5 \times S^5$ to yield a
conjectured equivalence between four-dimensional (4D) strongly coupled QCD at large $N_c$ and a
5D weakly interacting gauge theory coupled to gravity on a space $M_5$. Modifying the background geometry of the dual gravity theory induces confinement on the field theory side.  In the bottom-up approach,
one simply  cuts off $AdS_5$ at a finite radius, which
produces confinement and a
finite spectrum of resonances. In the top-down approach, the background geometry is that of $N_c$ $D4$-branes on an $S^1$ with boundary conditions that break supersymmetry. One can show that the resulting metric leads to an area law for Wilson lines \cite{Wittenconf}.
Chiral symmetry breaking is either produced by turning on a tachyonic field transforming in
the flavor bifundamental representation (hard-wall model, see e.g. \cite{Erlich:2005qh}),
via the appropriate infrared (IR) boundary conditions (Hirn-Sanz model of Ref.~\cite{Hirn:2005nr}), or
through the joining of $D8$-branes and $ \bar {D8}$-branes in the IR (Sakai-Sugimoto model of Ref.~\cite{Sakai:2004cn}).

In any of these  AdS/QCD models we conjecture that for every quantum operator ${\cal O}(x)$ in QCD,
there exists a corresponding bulk field $\phi(x,z)$, uniquely determined by the boundary condition $\phi(x,0) \equiv \phi_0(x) $
at the ultraviolet (UV) boundary of $M_5$.
Duality implies that the partition functions of the 4D gauge theory and 5D supergravity (or string theory) are equal.
Neglecting stringy corrections on the supergravity side, this amounts to a saddle-point evaluation of the
 bulk action $ S_{M_5}[\phi_0(x)]$ on the solution to the gravitational field equations:
\begin{align}
\langle \mbox{exp}(i\int d^4 x \phi_0(x){\cal O}(x))\rangle_{\rm QCD_4} = \mbox{exp}(iS_{\rm M_5}[\phi_0(x)]) \ .
\end{align}
The generating functional of the connected Green's functions, $ W_4[\phi_0(x)] $, equals
the 5D gravity action evaluated on the solution, so by varying both sides of the equation $ W_4[\phi_0(x)] = S_{M_5}[\phi_0(x)]$ with respect to $\phi_0(x)$ (and then setting these sources to zero), one can find the connected $n$-point Green's functions of the strongly coupled QCD.

As an instructive example, we explicitly compute the 2-point function for QCD vector currents in
the bottom-up approach
(for more details see Refs.~\cite{Erlich:2005qh,Grigoryan:2007vg}). First, we solve the gravity equations of motion, requiring
that the solution on the UV boundary coincide with the 4D source of the vector current. We then evaluate the
5D action on this solution to produce the generating functional and vary (twice)
with respect to the boundary sources.
It is convenient to work in $A_z = 0$ gauge and in terms of Fourier-transformed gauge fields, written as
$\tilde A_{\mu}(p,z) = \tilde{A}_{\mu}(p)\cV(p,z)$, where $\tilde A_{\mu}(p,z)$ and $\tilde{A}_{\mu}(p)$ are the Fourier transforms of $A_{\mu}(x,z)$
and the source $A_{\mu}(x)$ respectively \footnote{For simplicity, we ignore the flavor indices.}. The bulk-to-boundary propagator, $\cV(p,z)$, satisfies the
linearized supergravity equations of motion for a particular 4D momentum, $p$.
%
To quadratic order in the gauge fields, the 5D action can be written as:
\begin{align}
S_{{\rm AdS}_5}^{(2)} = -\frac{1}{2g^2_5}\int \frac{d^4p}{(2 \pi)^4}
\tilde{A}^{\mu}(p)\tilde{A}_{\mu}(p)\left[
\frac{1}{z}\partial_z \cV(p,z)\right]_{z=\epsilon} .
\end{align}
Varying this with respect to the boundary source gives the scalar part of the 2-point function:
\begin{align}
\label{sigmap2}
\Sigma(p^2) = -\frac{1}{g^2_5}\left(\frac{1}{z} \partial_z \cV(p,z)\right)_{z = \epsilon \rightarrow 0} \
\end{align}
which is in general defined by
\begin{equation}
\int~d^4x \ e^{i p \cdot x}\langle J_{\mu}(x)J_{\nu}(0) \rangle = \left( \eta_{\mu \nu} - \frac{{p_{\mu}p_{\nu}}}{p^2} \right)\Sigma(p^2) \ .
\end{equation}
Expanding the 2-point function $\Sigma(p^2)$ in the orthogonal basis of normalizable eigenfunctions generated by the 5D equations of motion,
we can see that the pole structure of the two-point function indeed corresponds to an infinite sum over resonances in which the poles  correspond to heavier and heavier vector mesons with masses  equal to the ratio of zeroes of the Bessel function $J_0$ to the IR cutoff $z_0$: $ M_{n} = \gamma_{0,n}/z_0 $, and the residues to the square of their decay constants. At large $p^2$ one finds
\begin{equation} \label{eq:veccor}
\Sigma(p^2) \sim  \frac{1}{2 g_5^2} p^2 \log(p^2) .
\end{equation}
Matching this result to perturbative QCD yields $g_5^2 = 12 \pi^2/N_c$ \cite{Erlich:2005qh}.

\subsection{The Hirn-Sanz Model}

The Hirn-Sanz model \cite{Hirn:2005nr} is very similar to the model introduced in \cite{Erlich:2005qh, DaRold:2005zs} and, like that model, incorporates fields dual to the operators in QCD which govern the structure of the low-energy theory, namely the chiral currents $J_{\mu L}^a$, $J_{\mu R}^a$. It differs in that chiral symmetry breaking is generated by infrared (IR) boundary conditions (BC) rather than through the expectation value of a tachyon field dual to the quark bilinear
$\bar q q$. The Hirn-Sanz model contains a pion field, but in contrast to the model of \cite{Erlich:2005qh, DaRold:2005zs} the pion cannot acquire a mass, and there are no higher Kaluza-Klein excitations of the pion field. These excitations are not relevant for our subsequent analysis, and excluding them makes it easier to disentangle the pion and axial-vector meson fields.

The model of Ref.~\cite{Hirn:2005nr} is based on the action
\begin{align}
S_{YM} &= -\frac{1}{4g^2_5}\int d^5 x \sqrt{g} \,\Tr\biggl[L_{MN}L^{MN} + R_{MN}R^{MN} \biggr] \ ,
\end{align}
in a background $AdS_5$ metric with an infrared cutoff:
\begin{align}\label{metric}
ds^2 =g_{MN}dx^Mdx^N = \frac{1}{z^2}\left(\eta_{\mu \nu}dx^{\mu}dx^{\nu} - dz^2\right) \ ,
\end{align}
where $ z \in (0,z_0) $, $ \eta_{\mu\nu} = {\rm Diag}(1,-1,-1,-1) $, $ \mu, \nu = (0,1,2,3) $, $ M, N =
(0,1,2,3,z) $,
\begin{align}
A_{MN} &= \partial_{M}A_{N} - \partial_{N}A_{M} - i [A_{M},A_{N}] \ ,
\end{align}
and $ A_{M} = T^a A^a_M $, $ A = \{L,R\} $ ($a = 0,1,2,3$). The gauge group is $U(2)$ with generators: $T^0= \textbf{1}/\sqrt{2N_f}$ and $T^i = \sigma^i/2$, where $\sigma^i$, $i=1,2,3$, are the Pauli matrices.

Under gauge transformations, the gauge fields  transform as
\begin{align}\label{gauge trans}
A_{M}(x,z) &\rightarrow g_AA_{M}g_A^{-1}(x,z) + i g_A\partial_{M}g_A^{-1}(x,z) \ ,
\end{align}
where $ g_{A}(x,z) \in U(2)_{A} $. At the UV boundary ($z=0$), the fields obey the BC $L_{\mu}(x,0) = \ell_{\mu}(x) $ and $R_{\mu}(x,0) = r_{\mu}(x) $, where $ \ell_{\mu}(x) $ and $ r_{\mu}(x) $ source the left and right 4D currents.
The vector $ V_{\mu} = (L_{\mu} + R_{\mu})/2 $ and axial-vector  $ A_{\mu} = (L_{\mu} - R_{\mu})/2 $ gauge fields are dual to
the vector and axial-vector currents of QCD, respectively.
Working in $ L_z(x,z) = R_z(x,z) = 0 $ gauge, one can write the vector field $ \hat{V}_{\mu}$
and the axial-vector field $\hat{A}_{\mu}$ in terms of boundary sources and dynamical fields as
\begin{align}
\hat{V}_{\mu}\left(x, z\right) &\equiv V_{\mu}\left(x,z\right) + \hat{V}_{\mu} \left(x,0\right) \ , \\ \nonumber
\hat{A}_{\mu}\left(x,z\right) &\equiv  A_{\mu}\left(x,z\right) + \alpha\left(z\right)\hat{A}_{\mu}\left(x,0\right)
\ ,
\end{align}
where the dynamical fields $ V_{\mu}\left(x,z\right) $ and $A_{\mu}\left(x,z\right)$ satisfy the following UV BC:
\begin{align}\label{BC1}
  V_{\mu}(x,0) &= 0 \ , \ \
  A_{\mu}(x,0) = 0 \ .
\end{align}
Chiral symmetry is broken by imposing different BC for $\hat{V}_\mu$ and $\hat{A}_\mu$ at the IR boundary ($z=z_0$).
The vector field obeys Neumann BC,
%
$ \partial_z V_{\mu}(x,z_0) = 0 $,
%
while both of the axial-vector fields $ A_{\mu}$ and $ \hat{A}_{\mu} $ satisfy Dirichlet BC
\begin{align}\label{BC1IRA}
A_{\mu}(x,z_0) = 0 \ , \ \ \hat{A}_{\mu}\left(x,z_0\right) = 0  \ .
\end{align}

As pointed out in Ref.~\cite{Hirn:2005nr} (and discussed in \cite{Grigoryan:2008cc}),
mixing between the pion and the axial resonances can be eliminated if the function $ \alpha \left(z\right) $
obeys the equation
\begin{eqnarray}
\partial_z \left( \sqrt{g} g^{\mu \nu} g^{zz} \partial_z \alpha(z)\right)  &=& 0 \ .  \label{eoma}
\end{eqnarray}
The BC on this field,
%
$ \alpha \left(0\right) = 1$, $\alpha \left(z_0\right) = 0$,
%
follow from Eqs.~(\ref{BC1}) and (\ref{BC1IRA}). As a result,
\begin{eqnarray}
  \alpha \left(z\right) & = & 1 - {z^2}/{z_0^2} \ .
\end{eqnarray}
The pion field in this model arises from the  chiral field
\begin{align}
U(x) = \xi_{R}(x)\xi_{L}^{-1}(x) \ ,
\end{align}
which is built from path-ordered Wilson lines:
\begin{align}\label{Wilson lines}
\xi_{L}(x) &= P \exp \left\{-i \int_0^{z_0} dz' L_z(x,z') \right\} \ , \\ \nonumber
\xi_{R}(x) &= P \exp \left\{-i \int_0^{z_0} dz' R_z(x,z') \right\}  \ .
\end{align}
The field $U(x)$ transforms under global chiral transformations in the same way as the chiral
field in the non-linear sigma model and can therefore be identified in the usual way with the pion
field through $U=e^{2 i \pi^a T^a/f_\pi}$.


The dynamical vector fields have the following decomposition
\begin{align}
V_{\mu}(x,z) = \sum^{\infty}_{n=1}V^{(n)}_{\mu}(x)\psi_{V,n}(z) \ ,
\end{align}
in terms  of the eigenfunctions  $ \psi_{V,n}(z) $ satisfying the equations of motion (EOM)
\begin{align}
\left[z^2 \partial_z^2 - z \partial_z + M^2_{V,n} z^2 \right]\psi_{V,n}(z) = 0 \ ,
\end{align}
with BC $ \psi_{V,n}(0) = \partial_z\psi_{V,n}(z_0) = 0 $. Here, e.g. the field $ V^{(1)}_{\mu}(x) = g_5\rho_{\mu}(x) $
describes the $\rho$-meson. The solution for $ \psi_{V,n}(z) $ is
\begin{equation}\label{eq:vecnormmode}
\psi_{V,n}(z) = \frac{\sqrt{2}}{z_0 J_1(\gamma_{0,n})}\, z J_1(M_{V,n} z) \ ,
\end{equation}
where $M_{V,n}$ is determined from $ J_0(M_{V,n} z_0) = 0 $ and, therefore, $ M_{V,n} = \gamma_{0,n}/z_0 $ (with $
J_0(\gamma_{0,n}) = 0 $). The value of $ z_0 = 1/(323 \ \MeV) $ is fixed from the experimental mass of the
$\rho$-meson $ M_{V,1} = 776 \ \MeV $. Eigenfunctions $ \psi_{V,n} $ are normalized as
\begin{align}
\int^{z_0}_0~ \, \frac{dz}{z} \, |\psi_{V,n}(z)|^2 = 1 \ .
\end{align}
The Fourier transform of the vector field is written as $ \tilde V_{\mu}(q,z) = \tilde{V}_{\mu}(q){\cal V} (q,z) $, where
$\tilde{V}_{\mu}(q)$  is the Fourier transform of the 4D field $V_{\mu} (x)$,  and ${\cal V} (q,z)$ is
the bulk-to-boundary propagator.  The latter satisfies the  EOM
\begin{equation}\label{JQzEOM}
z \, \partial_z\left(\frac{1}{z}\, \partial_z {\cal V}(q,z)\right) + q^2\, {\cal V}(q,z) = 0
\end{equation}
with BC ${\cal V}(q,0) = 1 $ and $ \partial_z{\cal V}(q,z_0) = 0 $. It can be also written as the sum
(for more details, see Ref.~\cite{Grigoryan:2007vg})
\begin{equation}\label{Jmeson}
{\cal V}(q,z) = -g_5\sum_{n = 1}^{\infty}\frac{f_{V,n} \psi_{V,n}(z)}{q^2 - M^2_{V,n} } \ ,
\end{equation}
where $ f_{V,n} $ is the  decay constant of $n^{\rm th}$ vector meson and
\begin{equation}
f_{V,n} = \frac{1}{g_5}\left[\frac{1}{z}\partial_z \psi_{V,n}(z)\right]_{z=0} = \frac{\sqrt{2}M_{V,n}}{g_5 z_0 J_1(\gamma_{0,n})} \ .
\end{equation}
%


The dynamical axial-vector fields can be written as:
\begin{align}
A_{\mu}(x,z) = \sum^{\infty}_{n=1}A^{(n)}_{\mu}(x)\psi_{A,n}(z) \ ,
\end{align}
where the functions $ \psi_{A,n}(z) $ satisfy the same EOM as $ \psi_{V,n}(z) $, but with different BC $ \psi_{A,n}(0) = \psi_{A,n}(z_0)
= 0 $. The solution can be written as:
\begin{align}\label{eq:avecnormmode}
\psi_{A,n}(z) &= \frac{\sqrt{2}}{z_0 |J_0(\gamma_{1,n})|}zJ_1(M_{A,n} z) \ ,
\end{align}
where $ M_{A,n} = \gamma_{1,n}/z_0$ with $ J_1(\gamma_{1,n}) = 0$. For $n=1$, using $ z_0 = 1/(323 \ \MeV) $, we get
$
M_{A,1} \equiv M_{a_1} = M_{f_1} \simeq 1237.64 \ \MeV \ .
$
Here in particular, the field $ A^{(1)}_{\mu}(x) = g_5 a_{1\mu}(x) $ describes the $a_1$-meson.


In the axial gauge, the axial-vector field with the dynamical fields turned off is  given by
\begin{eqnarray}
\hat{A}_{\mu} \left( x, z \right) &=& \alpha\left( z \right)\hat{A}_{\mu} \left( x, 0\right) =
\frac{i\alpha\left( z \right)}{2}\left\{\xi_L^{\dag}\partial_{\mu} \xi_L - \xi_R^{\dag}\partial_{\mu}\xi_R
\right\} \ .
\end{eqnarray}
Taking into account the definition for the Wilson lines $ \xi_{L,R}(x) $,
\begin{eqnarray}\label{defpifield}
\hat{A}^a_{\mu}(x,z) \approx \alpha(z)\partial_{\mu} \int^{z_0}_0 dz' \ A^a_z(x,z') \equiv
\alpha(z)(\partial_{\mu} \pi^a) \ .
\end{eqnarray}
The full axial-vector field in the axial gauge can be written as:
\begin{eqnarray}
\hat{A}_{\mu}(x,z) \simeq \alpha(z)\partial_{\mu} \pi(x) + \sum^{\infty}_{n=1}A^{(n)}_{\mu}(x)\psi^A_n(z)  \ .
\end{eqnarray}
%


To get a better idea on how the 5D gauge theory can reproduce the low energy behavior of QCD, we demonstrate the emergence of the order $ \cO(p^4) $ chiral Lagrangian from this holographic model.
To this end, we define the 4D field
\begin{align}
u_{\mu}\left(x\right) &= i \left\{ \xi_R^{\dag}\partial_{\mu}\xi_R - \xi_L^{\dag}\partial_{\mu}\xi_L \right\} \ .
\end{align}
One can check that if $ \ell_{\mu} = r_{\mu} = 0 $ and $ A_{\mu} = V_{\mu} = 0 $ (only chiral fields are turned on) then,
\begin{eqnarray}
  L_{z \mu} =  -\frac{1}{2}\left( \partial_z \alpha \right) \xi_L u_{\mu} \xi_L^{\dag} \ ,
  \ \ \ \ \
  R_{z \mu} = \frac{1}{2}\left( \partial_z \alpha \right) \xi_R u_{\mu} \xi_R^{\dag} \ ,
\end{eqnarray}
and
\begin{eqnarray}\nonumber
-\frac{1}{4g^2_5}\Tr\left(R^2_{\mu \nu} + L^2_{\mu \nu}\right)
= \frac{1}{32g^2_5}\left(1 - \alpha^2\right)^2\Tr\left[u_{\mu}, u_{\nu}\right]^2 \ .
\end{eqnarray}
Taking into account that
\begin{align}
  \xi_R u_{\mu} \xi_R^{\dag} &= - i U\partial_{\mu}U^{\dagger} \ , \ \ \ \ \xi_L u_{\mu} \xi_L^{\dag} = - i U^{\dagger}\partial_{\mu}U
\end{align}
and performing integration over $ z $, we get:
\begin{align}\label{Skyrme}
S_{YM} &= \int d^4 x \left\{\frac{f^2_{\pi}}{4}\Tr\left(\partial_{\mu}U^{\dagger}\partial^{\mu}U \right) +
\frac{1}{32e^2}\Tr\left[U^{\dagger}\partial_{\mu}U, U^{\dagger}\partial_{\nu}U\right]^2 \right\} \ ,
\end{align}
where
\begin{align}
f^2_{\pi} &= \frac{1}{g^2_5}\int^{z_0}_{0}\frac{dz}{z} \left(\partial_z \alpha\right)^2 = \frac{2}{g^2_5z^2_0} \ , \\[8pt]
\nonumber \frac{1}{e^2} &= \frac{1}{g^2_5}\int^{z_0}_{0}\frac{dz}{z}\left(1- \alpha^2\right)^2 =
\frac{11}{24g^2_5} \ .
\end{align}
This establishes the relation between 5D AdS/QCD and the 4D Skyrme model for two massless flavors.


We can use the formalism of AdS/QCD to (indirectly) compute couplings between the photon and various mesons. The photon couples to the hadronic part of the electromagnetic (EM) current $J^{{\rm EM}}_{\mu}$. Including a nonnormalizable mode of the gauge fields dual to this
current is equivalent to sourcing the current explicitly in the field theory Lagrangian. In
terms of the isosinglet $J_{\mu}^{\{I=0\}}$ and isovector $J_{\mu}^{\{I=1\}, 3}$ currents,
\begin{align}
J_{\mu}^{\{I=1\}, 3}&= \frac{1}{2}\left(\bar{u} \gamma_{\mu} u
- \bar{d} \gamma_{\mu} d \right) = \frac{1}{2}\bar{q}
\gamma_{\mu}\tau^3 q  \ ,
\\ \nonumber
J_{\mu}^{\{I=0\}} &= \frac{1}{2}\left (\bar{u} \gamma_{\mu} u + \bar{d} \gamma_{\mu} d \right)= \frac{1}{2}\bar{q}
\gamma_{\mu} \textbf{1}\, q \ ,
\end{align}
the hadronic EM current is
\begin{align}\label{eq:EMcurrent}
J^{{\rm EM}}_{\mu} = J_{\mu}^{\{I=1\}, 3}+ \frac{1}{3} \, J_{\mu}^{\{I=0\}} \ .
\end{align}
It is also useful to write the (hadronic) EM current as:
\begin{align}
J^{{\rm EM}}_{\mu} = \bar{q}\gamma_{\mu} Q_{\rm em}\, q = \bar{q}\gamma_{\mu} \left(I_3 + Y/2\right)q\ ,
\end{align}
where $ Q_{\rm em} = I_3 + Y/2 = {\rm diag}\{2/3, - 1/3\} $, $ I_3 = \tau^3/2 $ and $ Y = B + S = {\rm diag}\{1/3, 1/3\} $
(we are working in the two flavor case,  with zero strangeness).

As we have seen above, the nonnormalizable solution may be decomposed in terms of massive resonances.
This is a reflection of vector meson dominance.  The  EM current to which the photon couples is effectively carried by vector mesons, with the dominant contribution from $\rho^0$ and $\omega$. To produce a factor of the EM current from the holographic partition function, we must then apply the operator
\begin{align}
\frac{\delta}{\delta V^{{\rm em}}_{\mu}} \equiv \frac{\delta}{\delta V^3_{\mu}} + \frac{1}{3}\frac{\delta}{\delta \tilde{V}_{\mu}} \ ,
\end{align}
where $\tilde V_\mu$ is the $U(1)$ part of the vector current.

The bulk-to-boundary propagator for the vector fields $\cV(q,z)$ (described by the nonnormalizable mode)
can be written as in Eq.~(\ref{Jmeson}).
It can be shown that $\cV(0,z) = 1$: when the  photon inserted into the $q\bar{q}$ pair is on-shell, the  bulk-to-boundary propagator has a constant profile in the bulk. On the other hand, if the $q^2 = -Q^2 < 0 $,
this corresponds to the virtual photon with a nontrivial bulk profile given by
\begin{equation}
\label{JQz} \cV(Q,z) = {Qz}\left[K_1(Qz) + I_1(Qz) \frac{K_0(Qz_0)}{I_0(Qz_0)} \right] \ .
\end{equation}

One can employ similar current algebra arguments to incorporate hadronic weak interactions into the holographic model since the $W^\pm$ and $Z^0$ also couple
linearly to chiral currents.
This does not mean, however, that the holographic setup incorporates dynamical $W^{\pm}$ and $Z^0$ bosons. The interaction of these bosons with hadrons is realized through the chiral currents of QCD, just as the interaction of electromagnetism is realized through the vector currents. In this paper we will not explore the weak sector.
A more detailed discussion on incorporating weak interactions in a holographic setup can be found in  Ref.~\cite{Gazit:2008gz}.


\subsection{The Sakai-Sugimoto Model}

In contrast to the bottom-up model of Hirn and Sanz, the Sakai-Sugimoto model \cite{Sakai:2004cn,Sakai:2005yt} is a top-down
supergravity construction dual to  $SU(N_c)$ gauge theory with $N_f$ massless fermions in the fundamental representation. This construction generates the symmetries and degrees of freedom relevant
to QCD from a D-brane configuration in type IIA string theory. $N_f$ pairs of $D8$- and $\overline{D8}$-branes intersect
$N_c$ $D4$-branes as follows:
\begin{center}
\begin{tabular} {ccccccccccc}
 & $0$ & $1$& $2$& $3$ &$(4)$& $5$& $6$& $7$& $8$& $9$\\
$D4$ & x &x &x &x&x &&&&&\\
$D8-\overline{D8} \ \ $ & x& x& x& x & &x &x &x& x& x
\end{tabular}
\end{center}
The  $x_4\equiv \tau$ direction is wrapped on an $S^1$ of radius $M_{KK}^{-1}$ so $\tau\sim\tau+\delta\tau$ where $\delta\tau=2\pi M_{KK}^{-1}$. We break supersymmetry by imposing anti-periodic BC on the fermionic modes: while the $D$-brane gauge fields remain massless, their
fermionic superpartners acquire masses of order $M_{KK}$.

As the field theory living on the branes becomes strongly coupled ($\lambda = g^2_{YM}N_c \gg 1 $), and $N_c \rightarrow \infty$,
the stack of $D4$-branes is replaced by the supergravity background
\begin{align}\label{eq:D4sugra}
ds^2=\left(\frac{U}{R}\right)^{3/2}\left(\eta_{\mu\nu}dx^{\mu}dx^{\nu} + f(U)d\tau^2\right)
+ \left(\frac{R}{U}\right)^{3/2}\left(\frac{dU^2}{f(U)} + U^2 d\Omega_4^2 \right) \ , \\[5pt] \nonumber
e^\phi = g_s\left(\frac{U}{R}\right)^{3/4} \ , \ \quad F_4 = \frac{2\pi N_c}{V_4}\epsilon_4 \ , \ \quad f(U) \equiv 1 - \frac{U_{KK}^3}{U^3} \ .
\end{align}
The five directions transverse to the $D4$-brane are parametrized by a radial coordinate $U$ and
a unit $S^4$. Here $d\Omega_4^2$ is the metric on the unit $S^4$, which has volume form $\epsilon_4$ and volume $V_4=8\pi^2/3$. $U$ is bounded from below ($U\ge U_{KK}$) to avoid a conical singularity. The constant $R$ appearing
in the metric is $R^3=\pi g_sN_cl_s^3$.
In terms of $U_{KK}$ and $R$, $M_{KK}=3U_{KK}^{1/2}/(2R^{3/2})$, and the
4D Yang-Mills coupling has value $ g^2_{YM} = 2\pi M_{KK} g_s l_s $.

Keeping $N_f$ finite as $N_c \rightarrow \infty$, we treat the $D8$-branes as probes in the $D4$-brane background.
Extremizing  the $D8$-brane DBI action in this background, we find that the probes assume a nontrivial profile in the
$(\tau,U)$-plane. (Including the backreaction of the flavor branes is dual to including $N_f/N_c$ corrections in the field theory.)
At some $U=U_0$, the $D8$ and $\overline{D8}$ branes fuse into a single stack,
 a geometrical manifestation of chiral symmetry-breaking. At weak coupling, the
degrees of freedom at the $D4-D8$ and $D4-\overline{D8}$ intersections transform in the $(N_f,N_c)$ and
$(\overline{N_f},N_c)$ representations, while at strong coupling, these ``quarks'' and ``antiquarks'' are replaced by ``mesons'',
the $U(N_f)$ adjoint degrees of freedom on the fused $D8$ stack.

For the specific case studied in \cite{Sakai:2004cn} and \cite{Sakai:2005yt}, $U_0 = U_{KK}$. This corresponds to a
configuration where the $D8$ and $\overline{D8}$ are maximally separated in $\tau$ as $U\rightarrow\infty$. It is useful
to parametrize the $(U,\tau)$ plane in terms of
\begin{equation}
y = r\cos\theta \ , \ \ \  z = r\sin\theta
\end{equation}
where
\begin{equation}
r = U_{KK}\sqrt{\left( \frac{U}{U_{KK}}\right)^3-1} \ \quad \ \textrm{and} \ \quad \ \theta = \tau\frac{2\pi}{\delta\tau} \ .
\end{equation}
In these coordinates, the flavor brane profile is simply $y=0$, with $ z \in (-\infty,\infty) $.

The vector meson spectrum is generated by gauge and scalar fluctuations on the brane, which
obey the $D8$-brane DBI action:
\begin{align}\label{eq:GenSDBI}
S_{DBI} = -\mu_8 \int d^9x \ e^{-\phi} \Tr \sqrt{-\det\left(g_{MN} + 2\pi \alpha' F_{MN}\right)}
+ \mu_8 \int \sum  C_{p+1} \wedge \Tr e^{2\pi \alpha' F} \ ,
\end{align}
where
\begin{align}
\mu_8 \equiv \frac{2\pi}{(2\pi \ell_s)^9} \ , \ \ \ \ell^2_s = \alpha'
\end{align}
and $ \sum C_{p+1}$ is a formal sum of Ramond-Ramond fields of odd ranks that couple to $Dp$-branes (for $p = 0,2,4,6,8$).

The gauge field on the $D8$-brane has components
$ A_{\mu} $ ($\mu = 0,1,2,3$), $ A_z $ and $ A_{\alpha}$ ($\alpha = 5,6,7,8$, the coordinates on the $S^4$).
We are interested in states which are singlets under $SO(5)$, so we take $ A_{\alpha} = 0 $.
Assuming that the fluctuations ($ A_{\mu} $ and $ A_z $) do not depend on the $S^4$ coordinates,
the derivative expansion of the DBI action up to quadratic order is given by:
\begin{align}\label{eq:SDBI}
S_{YM} = \kappa\int d^4xdZ\Tr\left[-\frac{1}{2}K^{-1/3}F_{\mu\nu}F_{\rho\sigma}\eta^{\mu\rho}\eta^{\nu\sigma}+M_{KK}^2KF_{\mu Z}F_{\nu Z}\eta^{\mu\nu}\right] + \cO(F^3)
\end{align}
with the dimensionless coordinate $Z\in (-\infty, \infty)$ defined as $Z=z/U_{KK}$ and
\begin{align}
\kappa = \frac{3}{2}\frac{1}{(2\pi)^4 \ell_s^5g_s}\frac{U_{KK}^2}{M_{KK}^3}=\frac{N_c^2g_{YM}^2}{216\pi^3} \ \quad \ \textrm{and} \ \quad \  K(Z) = 1 + Z^2 \ .
\end{align}
%


On the other hand, in the normalization of Ref.~\cite{Sakai:2004cn}, the relevant term in the CS part of the action is:
\begin{align}\label{eq:CSDBIgeneral}
S_{CS} = \frac{1}{48\pi^3} \int_{D8} C_{3} \Tr F^3 =  \frac{1}{48\pi^3} \int_{D8} F_4 \ \omega_5(A) \ ,
\end{align}
where $ A = A_{\mu}dx^\mu + A_Z dZ $, $F = dA + iA \wedge A$, $ F_4 = dC_3 $ and $ \omega_5(A)$ is the CS 5-form:
\begin{align}\label{CS5form}
\omega_5(A) = \Tr\left(AF^2 + \frac{i}{2}A^3 F - \frac{1}{10}A^5\right) \ ,
\end{align}
such that $ d\omega_5(A) = \Tr F^3 $.
Since we assumed that gauge field fluctuations are independent of the $S^4$ coordinates,
\begin{align}
\int_{S^4}F_4 = 2\pi N_c \ ,
\end{align}
which is also expected from the second line of Eq.~(\ref{eq:D4sugra}).
As a result, the CS action can be written as \cite{Sakai:2004cn}:
\begin{align}\label{eq:CSDBI}
S_{CS} = \frac{N_c}{24\pi^2} \int_{M^4 \times R} \omega_5(A) \ .
\end{align}
%


To derive the 4D spectrum, we expand the dynamical gauge field $(A_\mu,A_Z)$ in terms of an orthogonal basis of eigenfunctions $\{\chi_n(Z)\}_{n\geq1}$ and $ \{\phi_n(Z)\}_{n\geq0} $ on $Z \in (-\infty,\infty )$
which vanish asymptotically as $ Z \rightarrow \pm\infty $:
\begin{align}
A_\mu(x^\mu,Z) &= \sum\limits_{n=1}^\infty B_\mu^{(n)}(x^\mu)\chi_n(Z) \ , \\ \nonumber
A_Z(x^\mu,Z) &= \varphi^{(0)}(x^\mu) \phi_0(Z) + \sum\limits_{n=1}^\infty\varphi^{(n)}(x^\mu )\phi_n(Z) \ ,
\end{align}
where $\chi_n(Z)$ satisfy the eigenvalue equation
\begin{align}\label{eq:eigenvalueSS}
\dt_Z\left( K\dt_Z\chi_n\right) = -\lambda_n^2K^{-1/3}\chi_n
\end{align}
and are normalized as
\begin{align}
\kappa\int dZ K^{-1/3}\chi_n\chi_m=\delta_{mn} \ ,
\end{align}
to give canonical kinetic terms for the $B_\mu^{(n)}$.
We now choose $\phi_n\propto\dt_Z\chi_n$ for $n\ge 1$ and $\phi_0(Z)=K^{-1}(\pi\kappa)^{-1/2}$ with normalization
\begin{align}
\kappa\int dZ K\phi_n\phi_m =\delta_{mn} \ .
\end{align}
Inserting the eigenfunction expansions into Eq~(\ref{eq:SDBI}) and integrating out over $Z$, we see that the massive modes of
$A_Z$, $\varphi^{(n)}$ can be absorbed into the $B_\mu^{(n)}$ as
\begin{align}
B_\mu^{(n)}\rightarrow  B_\mu^{(n)}-\frac{1}{\lambda_n}\dt_\mu\varphi^{(n)} \ .
\end{align}
This leaves a single tower of massive resonances $B_\mu^{(n)}$,
and a massless scalar, $\varphi^{(0)}$:
\begin{align}
S_{DBI} = -\int d^4x \Tr\left\{\frac{1}{2}\dt_\mu\varphi^{(0)}\dt^\mu\varphi^{(0)} + \sum_{n=1}^\infty\left[-\frac{1}{4}F^{(n)}_{\mu\nu}F^{(n)\mu\nu} + \frac{1}{2}m^2_n B_\mu^{(n)}B^{(n)\mu}\right]\right\} + \dots \ ,
\end{align}
where $ F^{(n)}_{\mu\nu}(x^\mu) \equiv  \dt_\mu B_\nu^{(n)} - \dt_\nu B_\mu^{(n)}$ and $ m_n \equiv \lambda_n M_{KK}$.

The tower of states $B_\mu^{(n)}$ includes both the vector and axial-vector resonances. The action and thus the eigenvalue equation (\ref{eq:eigenvalueSS}) are even under 5D parity, $(x^1,x^2,x^3,Z)\rightarrow (-x^1,-x^2,-x^3,-Z)$. The eigenfunctions therefore alternate in parity, with the lowest mode, $\chi_1$, being even, the second-lowest mode, $\chi_2$, being odd, etc. As a result, if $\chi_n$ is even (odd) under 5D parity, the corresponding $B_\mu^{(n)}$ is a vector (axial-vector) under 4D parity. The massless scalar $\varphi^{(0)}$ is a pseudoscalar identified with the pion field.

As in the model of Hirn and Sanz, the chiral field that yields the 4D Skyrme action  is identified with a Wilson line.
To obtain a finite 4D action for the normalizable modes, we implicitly required that $A_{M}(x^\mu,Z)$ should go to pure gauge
at the UV boundary ($ Z \rightarrow \pm \infty $).
As is discussed in Appendix A, the Chern-Simons part of the action is only defined up to boundary terms, which we assume
to vanish -- that is, we first write down the Chern-Simons action in a gauge where $A_M\rightarrow 0$ at the UV boundary.
We can again define Wilson lines
\begin{align}
\xi^{-1}_{\pm}(x^\mu) =P \exp\left\{-\int^{\pm \infty}_0 dZ' \ A_Z(x^\mu,Z') \right\} \
\end{align}
which define a transformation to $A_Z=0$ gauge.
Under this transformation, the boundary conditions at the UV boundaries are modified to
\begin{align}
A_{\mu}(x^\mu,Z) \rightarrow \xi_{\pm}(x^\mu)\partial_{\mu} \xi^{-1}_{\pm}(x^\mu)  \ , \ \ \ Z \rightarrow\pm \infty \ .
\end{align}
This allows us to expand the gauge field as follows:
\begin{align}
A_{\mu}(x^\mu,Z) = \alpha_{\mu}(x^\mu)\chi_0(Z) + \beta_{\mu}(x^\mu)  + \sum_{n=1}^{\infty}B^{(n)}_{\mu}(x^\mu)\chi_n(Z) \ ,
\end{align}
where
\begin{align}
\alpha_{\mu}(x^\mu) &\equiv \xi_{+}(x^\mu)\partial_{\mu}\xi^{-1}_{+}(x^\mu) - \xi_{-}(x^\mu)\partial_{\mu}\xi^{-1}_{-}(x^\mu)
\ , \\[5pt] \nonumber
\beta_{\mu}(x^\mu) &\equiv \frac{1}{2}\left(\xi_{+}(x^\mu)\partial_{\mu}\xi^{-1}_{+}(x^\mu) + \xi_{-}(x^\mu)\partial_{\mu}\xi^{-1}_{-}(x^\mu)\right) \ ,
\end{align}
and
\begin{align}
\chi_0(Z) \equiv \frac{1}{\pi}\arctan\left(Z\right) ~.
\end{align}
Using the residual gauge freedom to choose $\xi_+^\dagger=\xi_-=\sqrt{U(x)}$ and substituting this expansion into the action for $D8$-branes and integrating over $Z$, we obtain the Skyrme action, as in Eq.~(\ref{Skyrme}). In the Chern-Simons action it yields additional boundary terms treated in Appendix A, which are
identical to the boundary terms that result from making the same assumptions in the Hirn-Sanz model, and identifying $\xi_\pm^{(SS)}=\xi_{L,R}^{(HS)}$.


\subsection{Equivalence between Hirn-Sanz and Sakai-Sugimoto Models}

One of the purposes of this paper is to compare couplings derived in the prototypical bottom-up and top-down models of Hirn-Sanz and Sakai-Sugimoto.  In order to do so it is useful to make the similarities and differences in the models as clear as possible. Some elements of this similarity can be traced back to \cite{Son:2003et}.

The holographic coordinate in the Hirn-Sanz model lives on the interval $z\in (0,z_0)$,  with the UV boundary at $z=0$
and the IR boundary at $z=z_0$. The field content explicitly includes left- and right-handed gauge fields, $L_M$ and $R_M$. In Sakai-Sugimoto,
meanwhile, the coordinate $Z$ runs over a symmetric interval $(-\infty, \infty)$, (with the UV boundary at $Z \rightarrow\pm\infty$), and contains the single gauge
field we now call $\cA_M$. As discussed above the eigenfunction expansion of this $\cA_M$ splits into a piece which is symmetric under $Z\rightarrow -Z$
and a piece which is antisymmetric. In the 4d field theory, these correspond to vector and axial-vector modes, respectively. The basic idea is to use the symmetry properties under $Z \rightarrow -Z$ to map both the kinetic
and Chern-Simons terms into the interval $Z \in (0,+\infty)$. Working in ${\cal A}_Z=0$ gauge we thus
write
\begin{equation}\label{eq:SS_A_and_V}
\cA_\mu(x,Z) = V_\mu(x,Z) + A_\mu(x,Z) \equiv L_\mu(x,Z)~.
\end{equation}
Defining $R_\mu(x,Z) = V_\mu(x,Z)-A_\mu(x,Z)$ and using $L_\mu(x,-Z)=R_\mu(x,Z)$ and the fact that only  terms even
under $Z \rightarrow -Z$ survive integration over $Z\in (-\infty,\infty)$, we can rewrite the DBI action as
\begin{align}
S_{DBI}=\kappa \int_0^\infty dZ \int d^4x \Tr \biggl[-\frac{1}{2}K^{-1/3}\left(L^{2}_{\mu \nu}+R^{2}_{\mu \nu}\right)
+ M_{KK}^2 K \left(L^{2}_{\mu z}+R^{2}_{\mu z} \right) \biggr] + \cdots ~.
\end{align}
Up to the identification $2\kappa \leftrightarrow  1/g_5^2$ and a different choice of metric this gives
the Yang-Mills action similar to the one in the Hirn-Sanz model.
Now consider the Chern-Simons action of Sakai-Sugimoto, Eq. (\ref{eq:CSDBI}).
Using the expansion Eq.~(\ref{eq:SS_A_and_V}) we can write the CS action as
\begin{align}
S_{CS}=\int_{Z\in (-\infty,\infty)} \omega_5(V+A)
= \int_{Z\in (-\infty,0)}\omega_5(V+A) + \int_{Z\in (0,\infty )}\omega_5(V+A)
= \int_{Z\in (0,\infty)}\left( \omega_5(L)-\omega_5(R) \right)
\end{align}
after changing variables from $Z$ to $-Z$.

In the above we ignored the Wilson lines whose  values at the boundary are
dual to the pion modes but it is straightforward to include them --
it becomes quickly apparent that  identifying $\xi_\pm^{(SS)}=\xi_{L,R}^{(HS)}$ yields identical boundary terms in the Chern-Simons action of both models, and
thus Chern-Simons-induced couplings between pions and photons  are equal in the Hirn-Sanz and Sakai-Sugimoto models.

One can further bring out the similarity between the two models by changing coordinates so that the
UV and IR boundaries, and the IR boundary conditions, are mapped into each other. For example,
if one changes to coordinates $v=1/(Z+1)$ (where now $Z>0$) in the Sakai-Sugimoto model then $v$ runs from $0$ in the UV to
$1$ in the IR and the boundary conditions on the vector and axial-vector fields at $Z=0$ map to the same
boundary conditions at $v=1$, corresponding to the IR boundary conditions at $z/z_0=1$ in Hirn-Sanz.
Under this change of
variables the function appearing in front of the gauge field kinetic terms becomes $f(v)=(2 v^6-2v^5+v^4)^{-1/3}$
which should be compared with $w(z)=1/z$ in Hirn-Sanz. These two functions now have very similar profiles and asymptotics.

In order to compare the couplings determined below to experiment one must choose specific values for the parameters of the model. In the Hirn-Sanz model
$ z_0 = 1/(323 \ {\rm MeV})$ is fixed by fitting to the $\rho$ mass and $g_5 \simeq 4.91 $ by fitting to the experimental value for the pion decay constant ($ f^{\exp}_{\pi} = 92.4 \MeV $).  Alternatively, one could
choose the same value of $z_0$, but fit $g_5$ to the asymptotic value of the vector two-point function  Eq.~(\ref{eq:veccor}).
As in Ref.~\cite{Erlich:2005qh} this leads to $g_5= 2 \pi$ which leads to $f_\pi \sim 72~ {\rm MeV}$, in reasonable agreement with experiment  and consistent with the $1/N_c$ approximation we are using. In the
Sakai-Sugimoto model on the other hand the vector-vector two-point function has not been computed,
while fitting parameters to $f_\pi$ and the $\rho$ mass leads to $M_{KK}= 949 ~{\rm MeV}$ and $\kappa = .00745$.
In comparing results for the two models it seems most reasonable to fit both models to the same parameters,
hence in the calculations below and in the appendix we use $ z_0 = 1/(323 \ {\rm MeV})$ and $g_5 \simeq 4.91$ in the
Hirn-Sanz model. It should be kept in mind however that changing the value of $g_5$ by a factor of
roughly $1.2$ as is needed if we fit the vector two-point function instead leads to changes in the
normalization of wave functions which results in factors of $(1.2)^3 \sim 1.7$ in three-point couplings.


\section{Vertex Functions and Couplings from the Chern-Simons term in AdS/QCD}

We now apply the formalism described in the previous section to compute a set of couplings
that result from the presence of a Chern-Simons term in these two dual descriptions of QCD. For further
aspects of the connection between 5D Chern-Simons terms and 4D couplings see \cite{cth}.
Here we do computations in the Hirn-Sanz model.  Analogous calculations in the Sakai-Sugimoto model are straightforward and detailed results for both models can be found in Tables III and IV of Appendix A.
.
%
In the Hirn-Sanz model, the cubic part of the 5D CS action in the axial gauge $(L_z =R_z=0)$ is
\begin{align}\label{HSCSterm}
S^{\rm AdS}_{\rm CS}[L, R] = S^{(3)}_{\rm CS}[L]  - S^{(3)}_{\rm CS}[R] \ ,
\end{align}
where $ L=V+A$, and $R=V-A $ with  $L, R \in U(2)_{L(R)}$ and
\begin{align}
S^{(3)}_{\rm CS}[A] = \frac{N_c}{24\pi^2}\epsilon^{\mu\nu\rho\sigma}& {\rm Tr} \int d^4 x \int^{z_0}_0 dz
\left(\partial_z A_{\mu}\right)\biggl[\left(\partial_{\nu}A_{\rho}\right)A_{\sigma} + A_{\nu}\left(\partial_{\rho}A_{\sigma}\right)\biggr] \ .
\end{align}
%


\subsection*{Examples}

We will now compute a few specific couplings to illustrate the methods described earlier.  
\subsubsection{$\gamma \omega f_1$ and  $\omega \omega f_1$ Vertices}
The part of the 5D CS action which
contributes to the $\gamma\omega f_1$ coupling is:
\begin{align}\label{CSterm1}
S^{\rm AdS}_{\rm CS}[{\cal B}_L, {\cal B}_R] \supset \frac{N_c}{24\pi^2}\epsilon^{\mu\nu\rho\sigma} &\int d^4 x\, dz
\biggl(\partial_z \hat{V}_{\mu}\biggl[\left(\partial_{\nu}\hat{V}_{\rho}\right)\hat{A}_{\sigma} + \left(\partial_{\nu}\hat{A}_{\rho}\right)\hat{V}_{\sigma} \biggr] + \partial_z \hat{A}_{\mu}\left(\partial_{\nu}\hat{V}_{\rho}\right)\hat{V}_{\sigma}\biggr) \ .
\end{align}
We are interested in the following 3-point function:
\begin{align}
T_{\mu\nu\alpha}(q_1,q_2) &= \int d^4 x  d^4 y \ e^{iq_1x + iq_2y} \langle 0| \cT  J^{EM}_{\mu}(x)J^{\{I=0\}}_{\nu}(y)J^{\{I=0\}}_{A,\alpha}(0)|0\rangle \\ \nonumber
&\supset \frac{1}{3}\int d^4 x d^4 y \ e^{iq_1x + iq_2y} \langle 0| \cT J^{\{I=0\}}_{\mu}(x)J^{\{I=0\}}_{\nu}(y)J^{\{I=0\}}_{A,\alpha}(0)|0\rangle \ ,
\end{align}
where $ q_1 $ and $ q_2 $ are the momenta of the incoming photon and isosinglet vector meson respectively. To evaluate this correlator, we
vary the 5D CS action with respect to $ \hat{V}_{\mu}(p_1)$, $\hat{V}_{\nu}(p_2)$ and $\hat{A}_{\alpha}(p_3)$.
Note that at the end we have to divide this result by a factor of $3$, because of the form of the current in Eq.~(\ref{eq:EMcurrent}).
Factorizing the Fourier components of the fields as $ \hat{V}_{\mu}(q,z) = \tilde{V}_{\mu}(q)V(q,z) $ and $ \hat{A}_{\mu}(q,z) = \tilde{A}_{\mu}(q)A(q,z) $, so that
$V(q,0) = A(q,0) = 1$, we find
\begin{align}
T^{\mu\nu\alpha}(p_1,p_2) &= \frac{N_c}{24\pi^2}\epsilon^{\mu\nu\alpha\beta}K_{\beta}(p_1,p_2) \ i(2\pi)^4\delta^{(4)}(p_1 + p_2 + p_3) \ ,
\end{align}
where, using $p_1 + p_2 + p_3 = 0 $ and defining $ q = -(p_1+p_2)$,
\begin{align}\label{CSVertexGen}
K_{\beta}(p_1,p_2) \equiv \frac{1}{3}(p_1 -p_2)_{\beta}&\left[V(p_1,z)V(p_2,z)A(q,z)\right]^{z_0}_0 \\ \nonumber
&+ \int^{z_0}_0 dz \left[p_{2\beta}V(p_2,z)\partial_zV(p_1,z) - p_{1\beta}V(p_1,z)\partial_zV(p_2,z) \right]A(q,z)   \ .
\end{align}
Placing the photon and $f_1$ meson on shell we have $p^2_1 = 0$, $q^2 = M^2_{f_1}$, $ V(p_1,z) \rightarrow 1 $ and
$ A(q,z) \rightarrow g_5 \psi_{f_1}(z) $ we obtain:
\begin{align}\label{VertFunc}
\frac{N_c}{24\pi^2} \ K_{\beta}(p_1,p_2) \ \rightarrow \ -\frac{1}{4\pi} \  p_{1\beta}\int^{z_0}_0 dz \ \partial_zV(p_2,z) \psi_{f_1}(z) \ .
\end{align}
To find the $g_{\gamma \omega f_1}$ coupling, we place the $\omega$ on-shell.
Substituting $V(p_2,z)$ with $ g_5 \psi_{\omega}$ in Eq.~(\ref{VertFunc}), we get:
\begin{align}
e g_{\gamma \omega f_1} =  \ e\int^{z_0}_0 dz \ \psi_{\rho}(z)\partial_z\psi_{a_1}(z) \simeq -0.34~e \simeq -0.10 \ ,
\end{align}
where $e = \sqrt{4\pi \alpha_{\rm em}}$ is the electromagnetic charge. In the framework of the HS model \cite{Hirn:2005nr},
we used the wavefunctions from Eqs. (\ref{eq:vecnormmode}) and (\ref{eq:avecnormmode}) for the light (axial)vector mesons.

Similarly, the part of the 5D CS action which contributes to the $\omega\omega f_1$ coupling is also given by Eq.~(\ref{CSterm1}). The vertex function is therefore the same as  in Eq.~(\ref{CSVertexGen}) but multiplied by a factor of $3$. Taking into account that $ \hat{V}_{\mu}= g_5 \omega_{\mu}(x)\psi_{\rho}(z) $ and
$ \hat{A}_{\mu}= g_5 f_{1\mu}(x)\psi_{a_1}(z) $ (assuming $\psi_{f_1}=\psi_{a_1}$) we find:
\begin{align}\label{WW1AdS}
g_{\omega\omega f_1} &= 3g^3_5\frac{N_c}{24\pi^2} \int^{z_0}_0 dz \ \psi_{a_1}\psi_{\rho}\partial_z \psi_{\rho}  \simeq -4.09  \ .
\end{align}
%


\subsubsection{$\gamma \rho^0 f_1$ Vertex}

The part of the 5D CS action which contributes to the $\gamma\rho^0 f_1$ coupling is
\begin{align}
S^{\rm AdS}_{\rm CS}[L,R] \supset \frac{N_c}{48\pi^2}\epsilon^{\mu\nu\rho\sigma} &\int d^4 x\, dz
\biggl(\partial_z V^a_{\mu}\biggl[\left(\partial_{\nu}V^a_{\rho}\right) {\hat A}_{\sigma} + \left(\partial_{\nu}{\hat A}_{\rho}\right)V^a_{\sigma} \biggr] + \partial_z \hat{A}_{\mu}\left(\partial_{\nu}V^a_{\rho}\right)V^a_{\sigma}\biggr) \ ,
\end{align}
for $U(1)$-valued ${\hat A}_\mu$.
Varying the action gives
\begin{align}
\tilde{T}^{\mu\nu\alpha}(p_1,p_2) &= \frac{N_c}{24\pi^2}\epsilon^{\mu\nu\alpha\beta}\tilde{K}_{\beta}(p_1,p_2) \ i(2\pi)^4\delta^{(4)}(p_1 + p_2 + p_3) \ ,
\end{align}
where as before we have used $ p_1 + p_2 + p_3 = 0 $ and defined $ q = -(p_1+p_2)$. Using
$ p^2_1 = 0 $ we then have
\begin{align}
\tilde{K}_{\beta}(p_1,p_2) \equiv (p_1 -p_2)_{\beta}&\left[V(p_1,z)V(p_2,z)A(q,z)\right]^{z_0}_0 + 3p_{1\beta}\int^{z_0}_0 dz \partial_zV(p_2,z)A(q,z)   \ .
\end{align}
which yields
\begin{align}
eg_{\gamma\rho^0 f_1} &= 3eg_{\gamma\omega f_1} \simeq -0.31  \ .
\end{align}

Further couplings can be found in Tables III and IV of Appendix A. In Appendix C we compute the decay rate for
$f_1 \rightarrow \rho^0 + \gamma$ using the above coupling and find reasonable agreement with the measured rate.


\section{Photoproduction}

In this section we use the anomalous couplings derived above to study  the exclusive meson ($M$)
photoproduction process ($\gamma(p_1) + N(q_1) \rightarrow M(p_2) + N(q_2) $).
Photoproduction of vector mesons with the same quantum numbers as the photon (e.g. $\rho^0$, $\omega$, $\phi$) is
well-studied experimentally and theoretically, using the standard tools of Vector Meson Dominance (VMD)
and Regge phenomenology, see e.g. Refs.~\cite{Collins:1977jy} and \cite{photorefs}.
These processes receive contributions from the exchange of particles or families of particles with vacuum quantum numbers, and are therefore thought to be dominated by Pomeron exchange in the Regge limit (large $ s = (p_1 + q_1)^2 $ and fixed $ t = (p_1 - p_2)^2 $). At small $|t|$ this involves
exchange of the ``soft Pomeron,'' while with increasing $|t|$ one encounters the complicated issue of the transition to the ``hard Pomeron" and its relation to perturbative QCD.

Here we focus first on small $ s $, where single particle exchange should be a good approximation, and then on the large $s$, small $|t|$ regime of standard Regge theory. The relatively new element is that we study processes where the exchanged objects do not carry vacuum quantum numbers. This allows us to probe the anomalous couplings derived from dual models. We will focus on photoproduction of $f_1$ meson, as this process is the cleanest from a theoretical standpoint, and is most likely to have a clear experimental signature. We also briefly consider  the contribution of the exchange of $f_1$ mesons and their associated Regge trajectory to the photoproduction of $\rho^0$ and $\omega$ mesons.

\subsection{$f_1$ photoproduction}

In this section we study the photoproduction of  the $f_1(1285)$ isosinglet axial-vector meson.
%
One can see from Table I that in order to preserve  charge conjugation symmetry  the exchanged particles must have $ C = - 1 $. The lightest mesons with $ C = - 1 $ are $\omega$ and $\rho$. One can check that neither $\pi^0$ or $\sigma$ contribute to  photoproduction. The Pomeron as well as the tensor mesons $f_2(1270)$, $a_2(1320)$ and so on also have $C=+1$ and so do not
contribute to this process.


\begin{center}
\begin{tabular}[t]{|l|c|c|c|c|c|}
\multicolumn{6}{c}{TABLE I: Spin-1 states.}\\  \hline
  $ \ V/A \ $      & \ $ P $ \ & \  $ C $ \ & \ $ G $ \ & \  $ I $ \   & \ $ \ \xi \ $    \\ \hline
  $ \ \gamma \ $ \ & \ $-1 $ \  & \ $-1$ \  & \  \  \   & \  $ \  $ \  & \ \ \ \     \\
  $ \ \omega \ $ \ & \ $-1 $ \  & \ $-1$ \  & \ $-1$ \  & \  $ \ 0$ \  & \ $-1 \ $   \\
  $ \ \rho   \ $ \ & \ $-1 $ \  & \ $-1$ \  & \ $+1$ \  & \  $ \ 1$ \  & \ $-1 \ $  \\
  $ \ f_1 \ $  \   & \ $+1 $ \  & \ $+1$ \  & \ $+1$ \  & \  $ \ 0$ \  & \ $-1 \ $    \\
  \hline
\end{tabular}
\end{center}

\begin{center}
\begin{tabular}[t]{|l|c|c|c|c|}
\multicolumn{5}{c}{TABLE II: Some useful parameters for $\rho$, $\omega$ and $f_1$ mesons.
} \\  \hline
  $ \ \ M \ $         & \ $ m_M \ (\MeV) $ \ & \ $ g_{MNN} $ \ & \ $ g_{\gamma M f_1} $ \ & \ $\alpha_M(t) \ $  \\ \hline
  $ \ \ \omega \ $ \ & \ $ 782.6 $ \        &   \ $ 9   $ \   & \  $ \ -0.33 \ $ \      & \ $0.44 + 0.9t/\GeV^2 \ $   \\
  $ \ \ \rho   \ $ \ & \ $ 768.5 $ \        &   \ $ 2.4 $ \   & \  $ \ -0.99 \ $ \      & \ $0.55 + 0.8t/\GeV^2 \ $ \\
  $ \ \ f_1 \ $  \   & \ $ 1281.8$ \        &   \ $ 2.5 $ \   & \  $ \   \   $        & \ $ \alpha_{f_1}(0) +  \alpha'_{f_1}t \ $  \\
  \hline
\end{tabular}
\end{center}

In principle, $f_1$ photoproduction can occur with the exchange of a virtual photon.
However, direct calculations show that at small $|t|$ and large $s$, the contribution from this process is insignificant
compared to the process involving vector meson exchange.

Finally, there are also contributions from $u$- and $s$- nucleon exchange channels (see Fig.~\ref{Fig3} of Appendix B).
As  discussed in Appendix B, for $ s \gg M^2_N $ and fixed $t$, the contribution from these channels is small. We will ignore it in our analysis
(the same holds true for Regge trajectory exchange).
In addition, the polarization structure resulting from nucleon exchange is  quite different
from that in the $t$-channel exchange mechanism. The $t$-channel exchange amplitude involves the antisymmetric Levi-Civita tensor
(see e.g. Eq.~(\ref{eq:Mforf1})) while the nucleon contributions do not: this  may also allow them
to be separated out in experiments.
%


\subsection{Single Particle Exchange: Small $s$ Amplitudes}

Single particle exchange should dominate for energies below or slightly above the threshold.
The contribution of  vector meson exchange  to the amplitude for $f_1$ photoproduction
can be written as:
\begin{align}\label{eq:Mforf1}
\cM_{\rm sp}(\gamma p \rightarrow f_1 p) = i e \ & \epsilon^{\mu\nu\alpha\beta} \ \epsilon^*_{\mu}(\gamma)\epsilon_{\nu}(f_1) \ p_{1\beta} \ K_{\gamma V f_1}(t) \ \left(g_{\alpha\delta} - \frac{q_{\alpha}q_{\delta}}{M^2_V} \right) \ \frac{1}{t- M^2_V} \\[7pt] \nonumber  & \ \times \ g_{VNN} \ F_{V}(s,u)  \ \bar{u}_{s'}(q_2)\left[\gamma^{\delta} + i \kappa_V \sigma^{\delta \lambda}\frac{q_{\lambda}}{2M_N}\right]u_s(q_1) \ ,
\end{align}
where $ q = p_1 - p_2 $, $ t = q^2 = -Q^2 < 0 $, $\epsilon_{\mu}(\gamma)$ and $\epsilon_{\nu}(f_1)$ are the polarization vectors
of the  initial photon and final $f_1$
meson, the subscripts $s$ and $s'$ denote polarizations of initial and final nucleons, and $K_{\gamma V f_1}(t)$ is a vertex function,
related to the one obtained from holographic QCD in the previous section. It will be convenient to write the vertex function as:
$K_{\gamma V f_1}(t) = g_{\gamma V f_1}F_{\gamma V f_1}(t) $,
where $F_{\gamma V f_1}(t)$ is such that $F_{\gamma V f_1}(M^2_V) = 1 $. For small values of $|t|$, the exchange of the lightest vector mesons ($\rho$ or $\omega$) is dominant: from now on we will neglect any contribution from the daughter trajectories, setting $F_{\gamma Vf_1} = 1$.

The coupling $g_{VNN}$ describes the interaction of the vector meson with the nucleon, see Table II.
The Lagrangian governing this interaction can be written as
\begin{align}
\cL_{VNN}  = -g_{VNN} \ \bar{\psi}_N\left[\gamma^{\mu}V_{\mu} - \frac{\kappa_V}{2M_N}\sigma_{\mu\nu}\partial^{\nu}V^{\mu}\right]\psi_N \ ,
\end{align}
where $V_{\mu}$, $\psi_N$ are  the vector meson field and nucleon field respectively.
The nucleon form factor $F_{V}(s,u)$, corresponding to the $VNN$ vertex,
is taken from Ref.~\cite{Oh:2000pr} (see also Ref.~\cite{Haberzettl:1998eq}):
\begin{align}
F_{V}(s,u) = \frac{1}{2}\left[\frac{\Lambda^4_{VNN}}{\Lambda^4_{VNN} + (s-M^2_N)^2} + \frac{\Lambda^4_{VNN}}{\Lambda^4_{VNN} + (u-M^2_N)^2} \right] \ ,
\end{align}
where $\Lambda_{VNN} = 0.8 \ \GeV$.

Using the antisymmetry of  $\sigma^{\delta \lambda}$ and the Dirac equation $\sla{p}u(p) = M_Nu(p) $ we can write
\begin{align}
\left(g_{\alpha\delta} - \frac{q_{\alpha}q_{\delta}}{M^2_V} \right) \ \bar{u}_{s'}(q_2)\left[\gamma^{\delta} + i \kappa_V \sigma^{\delta \lambda}\frac{q_{\lambda}}{2M_N}\right]u_s(q_1) = \bar{u}_{s'}(q_2)\left[\gamma^{\alpha} + i \kappa_V \sigma^{\alpha \lambda}\frac{q_{\lambda}}{2M_N}\right]u_s(q_1) \ ,
\end{align}
where we have used $ q = q_2 - q_1$ and $ \bar{u}_{s'}(q_2)(\sla{q_2} - \sla{q_1})u_s(q_1) = 0 $.
From now on, we will ignore the contribution from the term proportional to $\kappa_V$,
as it does not play a significant role in our later discussions.
Therefore, the scattering amplitude can be written as:
\begin{align}
\cM_{\rm sp}(\gamma p \rightarrow f_1 p) = i e \ g_{VNN} \ g_{\gamma V f_1} \ & \epsilon^{\mu\nu\alpha\beta} \ \epsilon^*_{\mu}(\gamma)\epsilon_{\nu}(f_1) \ p_{1\beta} \   \ \frac{1}{(t - M^2_V)} \ F_{V}(s,u) \ \bar{u}_{s'}(q_2)\gamma^{\alpha} u_s(q_1) \ .
\end{align}
From Table II, we have $ g_{\gamma \rho^0 f_1} = 3g_{\gamma \omega f_1}$, $g_{\omega NN} \simeq 9 $, $ g_{\rho NN} \simeq 2.4 $
\footnote{Within the framework of the Sakai-Sugimoto model, Ref.~\cite{Hong:2007ay} derives the following values for these couplings:
$ g_{\omega NN} \simeq 12.6 $, $ g_{\rho NN} \simeq 3.6 $.},
which gives the ratio
\begin{align}
\cR \equiv \frac{g_{\omega NN} \ g_{\gamma \omega f_1}}{g_{\rho NN} \ g_{\gamma \rho f_1}}
= \frac{g_{\omega NN}}{3g_{\rho NN}}  \simeq 1.25 \ .
\end{align}
As a result, the sum of the amplitudes can be written as:
\begin{align}
\cM^{\rm tot}_{\rm sp}(\gamma p \rightarrow f_1 p) = \cM_{\rho-{\rm exch}}(\gamma p \rightarrow f_1 p) + \cM_{\omega-{\rm exch}}(\gamma p \rightarrow f_1 p) \simeq  r \ \cM_{\rho-{\rm exch}}(\gamma p \rightarrow f_1 p) \ ,
\end{align}
where $ r \equiv 1+ \cR = 2.25 $.



The differential cross section for $f_1$ photoproduction can be written as:
\begin{align}
\frac{d\sigma_V}{dt} = \frac{1}{16\pi (s-M^2_N)^2}|\cM^{\rm tot}_{\rm sp}(\gamma p \rightarrow f_1 p)|^2 \ .
\end{align}
If we do not keep track of the polarization structure then we should average over initial photon
polarizations and sum over final $f_1$ polarizations using
\begin{align}\label{unpol}
\sum_{\rm pol} [\epsilon^*_{\mu}\epsilon_{\mu'}]_{\gamma} = -g_{\mu\mu'} \ ,  \ \ \  \sum_{\rm pol}[\epsilon^*_{\nu'}\epsilon_{\nu}]_{f_1} = -g_{\nu'\nu} + \frac{p_{2\nu'}p_{2\nu}}{M^2_{f_1}} 
\end{align}
which leads to
\begin{align}\label{eq:SPcrosssec}
\frac{d\sigma_V}{dt} &= \frac{A_V F^2_{V}(s,u)}{8(t - M^2_{\rho})^2}\Phi_s(s,t) \ ,
\end{align}
where $ A_V \equiv r^2 \alpha_{\rm em} \ g_{\rho NN}^2 g_{\gamma \rho f_1}^2 $ and
\begin{align}\label{eq:Psform}
\Phi_s(s,t) = \frac{M^2_N- u}{s - M^2_N}
+ \frac{(M^2_{f_1} - t)}{2M^2_{f_1}}\left[\frac{(M^2_N - u)(M^2_N + M^2_{f_1} - u)}{(s - M^2_N)^2} + \frac{(s - M^2_N - M^2_{f_1})}{(s - M^2_N)} - 2M^2_N\frac{(M^2_{f_1} - t)}{(s - M^2_N)}\right] \ .
\end{align}
%


Now let us  briefly consider polarized $f_1$ photoproduction.  We assume  for simplicity that only the photon and $f_1$ meson are in polarized states while the initial and final protons are  unpolarized. (The generalization to include polarized protons is straightforward.)
In a general treatment  the products $[\epsilon^*_{\mu}\epsilon_{\mu'}]_{\gamma}$ and $[\epsilon^*_{\mu}\epsilon_{\mu'}]_{f_1}$ that arise in  $|\cM^{\rm tot}_{\rm sp}(\gamma p \rightarrow f_1 p)|^2$ would be
replaced by  density matrices  $\rho^{\gamma,f_1}_{\mu'\mu} $ and one would calculate various spin observables and helicity amplitudes, see e.g. Ref.~\cite{Savkli:1995vf}.

Here we will be content to note that the presence of the Levi-Civita tensor in the invariant matrix element
leads to the following characterization of the polarization structure. Let us work in the rest frame of the
target proton in which case $\bar u (q_2)\gamma^\alpha u(q_1)$ is nonzero only for $\alpha=0$.
The invariant amplitude Eq.~(\ref{eq:Mforf1}) is thus proportional to the volume of a parallelpiped spanned by  the photon momentum $\vec p_1$
and the spatial components of the photon and $f_1$ polarization vectors $ \vec \epsilon^*(\gamma)$, $ \vec \epsilon(f_1)$. It reaches its maximal value when these
form a mutually orthogonal set of three-vectors.
%


\subsection{Reggeon Exchange: large $s$ and small $t$}

To analyze $f_1$ photoproduction in the Regge region of large $s$ and fixed $t$  we adopt the standard prescription whereby the Feynman single-particle propagator  is replaced by the Reggeized propagator
corresponding to exchange of the particles Regge trajectory (see e.g. \cite{Collins:1977jy,Irving:1977ea,Guidal:1997hy})
\begin{align}\label{Reggeprop}
\frac{1}{t - M^2_V} \ \rightarrow \   \frac{\pi \ \alpha'_V}{2 \ \Gamma( \alpha_V(t))} \ \eta(t)\left(\frac{s}{s_0}\right)^{\alpha_V(t)-1} \equiv \cP_R(s,t) \ ,
\end{align}
where $s_0 = 1 \ \GeV^2$ and
\begin{align}
\eta(t) = - \frac{1+\xi e^{-i\pi\alpha(t)}}{\sin \pi \alpha(t)} = i - \tan\left(\frac{\pi \alpha(t)}{2}\right) 
\end{align}
is a signature factor (with $\xi = -1$ for either $\omega$ or $\rho$, see Table I).
The gamma function $\Gamma( \alpha_V(t))$ in Eq.~(\ref{Reggeprop}) suppresses poles of the propagator in the unphysical region (so-called nonsense poles).
For vector and tensor mesons, the Regge trajectories  are all approximately equal (a fact known as exchange degeneracy) and may be written as $\alpha_V(t) \simeq 0.5 + 0.9 \ t $ ($\rho$, $\omega$, $a_2$, $f$, etc.), see, e.g., Ref.~\cite{Collins:1977jy} (more precise trajectories for some of the mesons can be found in Table II). One interesting consequence of Eq.~(\ref{Reggeprop}) is the existence of zeroes (called EXD zeroes)
at $\alpha_V(t) = -2n$ with $n$ a non-negative integer. The first of these zeroes occurs at $ t = -\alpha_V(0)/\alpha'_V \in [-0.5,-0.7]$ depending on which precise trajectory  contributes to a particular process. These zeroes are thought to be responsible for dips in differential cross sections, see for example
\cite{Barnes:1976ek}.

Applying the above mentioned prescription in Eq.(\ref{Reggeprop}),
the contribution to the $f_1$ photoproduction amplitude from both $\rho$ and $\omega$ meson Regge trajectory exchanges can be written as:
\begin{align}\label{ReggeEx}
\cM_{\rm R} = i \ r \ e \ g_{\rho NN} \ g_{\gamma \rho f_1} \ \epsilon^{\mu\nu\alpha\beta} \ \epsilon^*_{\mu}(\gamma)\epsilon_{\nu}(f_1) \ p_{1\beta}  \ \cP_R(s,t) \ F(t) \ \bar{u}(q_2) \gamma^{\alpha}u(q_1) \ .
\end{align}
In Eq.~(\ref{ReggeEx}), $F(t)$ is a form factor governing the coupling of vector mesons to the proton.  Such form factors are usually  well-approximated by a dipole form,
\begin{align}
F(t) =  \left(\frac{1}{1- t/M_d^2}\right)^2  \ .
\end{align}
We will assume this form in what follows.  To get explicit results, we will take $M_d^2 = 0.71~ {\rm GeV}^2$
which is the dipole mass appearing in the electromagnetic form factor often used in Regge theory. It should be noted, however,
that the actual dipole mass for this process may be somewhat different.


In the unpolarized case, for large $s$ and small $ |t| $, we have:
\begin{align}
\frac{d\sigma_R}{dt} \simeq \pi^2 A_V F^2(t) \ \frac{\left(1 - \frac{t}{2M^2_{f_1}}\right) \ \alpha^{'2}_V}{16 \ \Gamma^2( \alpha_V(t)) \ \cos^2\left(\frac{\pi \alpha_V(t)}{2}\right)} \ \left(\frac{s}{s_0}\right)^{2(\alpha_V(t)-1)}  \ ,
\end{align}
where $ \alpha_V(t) \simeq 0.5 + 0.9 t $.

\begin{figure}[h]
\includegraphics[width=5.5in]{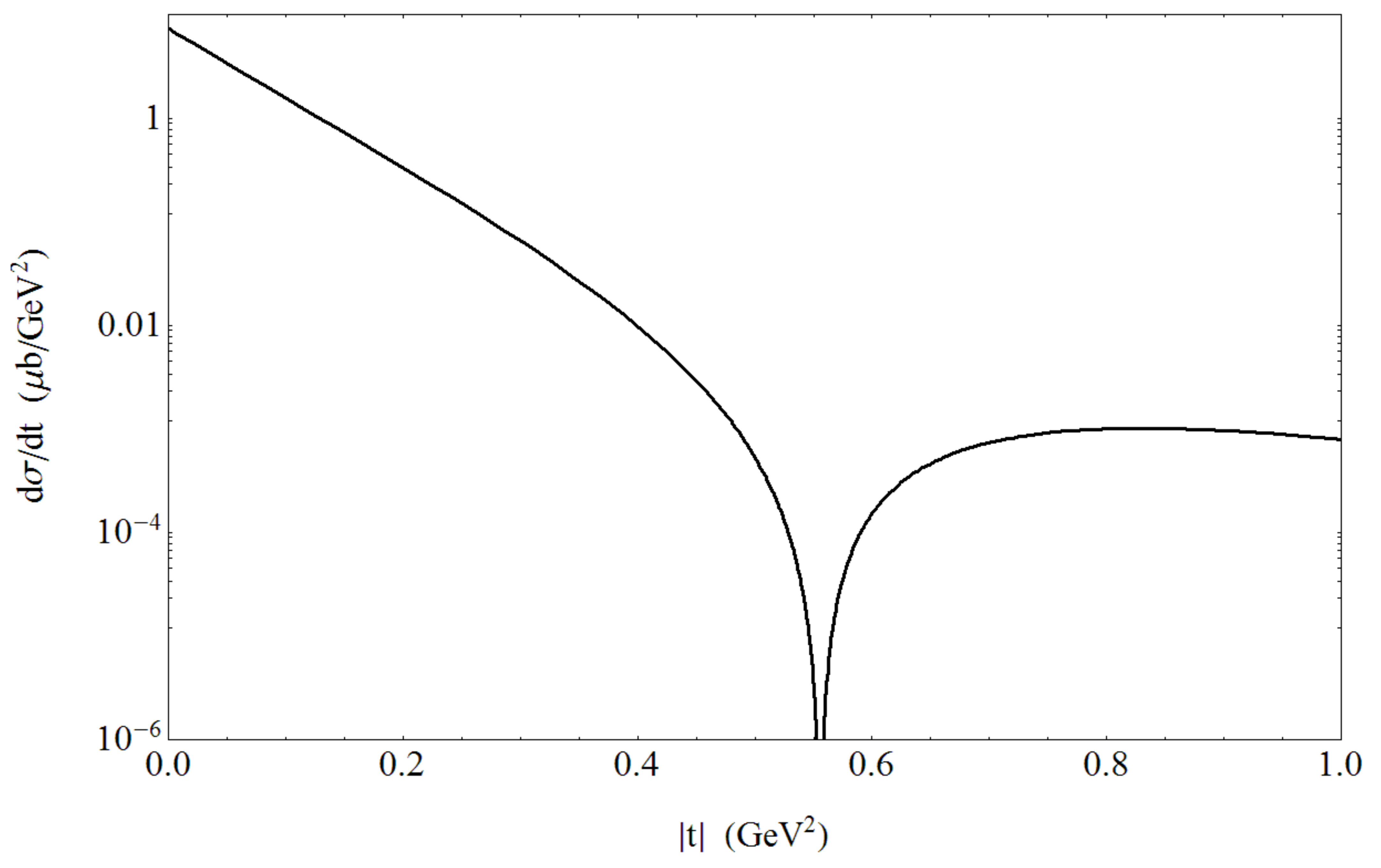}
\caption{\label{FigAB} The contribution of $\omega$ and $\rho$-meson Regge trajectories to the differential cross section $d\sigma_R/dt$ for $f_1$ photoproduction for $|t|\le 1 {\rm GeV}^2$ at $s = 10 \ \GeV^2$ ($E_{\gamma} \simeq  4.6 \ \GeV$ in lab frame).}
\end{figure}

The differential cross-section for $f_1$ photoproduction is shown in Fig.~\ref{FigAB}.
As expected, it exhibits a strongly collimated peak at forward angles corresponding to $t$-channel meson exchange. In the physical region, where $t \leq 0$, the differential cross section scales approximately as  $s^{-1-1.8|t|}$. Fig.~\ref{FigAB} also exhibits an EXD zero at $|t| \simeq 0.55 \ \GeV^2 $.

It is instructive to compare this contribution to the one from the nucleon's $s$- and $u$-channel Regge trajectory exchange.
Since the Regge trajectory for the nucleon is $ \alpha_N(t) \simeq - 0.3 + 0.9 \ t $ (the values for the slope and intercept are taken from the Ref.~\cite{Collins:1977jy}), we expect that at large $s$, the differential cross section should decrease  like $ s^{-2.6}$. This suggests that for large $s$ exchange of  the $\omega$ and $\rho$  Regge trajectories is dominant.


\subsection{$\omega$ ($\rho$) photoproduction}

For the $\omega$ or $\rho^0$ meson photoproduction at high center of mass energies, the Regge trajectory exchange (from $\pi$, $\sigma$ and $f_1$ mesons) must also be taken into account. The single particle contributions that also include tensor mesons are considered in Ref.~\cite{Oh:2003aw}. We will only be interested in the contribution from the Regge trajectory corresponding to $ f_1$ meson. This trajectory should be dominant, since the Regge trajectory for the pion is $ \alpha_{\pi}(t) \simeq 0.0 + 0.8 \ t $, therefore, the differential cross section will decrease as $ s^{-2}$.


The contribution of the single-particle exchange to the $\omega$ photoproduction amplitude
can be written as:
\begin{align}
\cM_{\rm sp} = i \ e \ g_{f_1NN} \ g_{\gamma \omega f_1} & \epsilon^{\mu\nu\alpha\beta} \ \epsilon^*_{\mu}(\gamma)\epsilon_{\nu}(\omega) \ p_{1\beta} \ \frac{1}{t - M^2_{f_1}} \ F_{f_1NN}(t) \ \bar{u}_{s'}(q_2) \gamma^{\alpha}\gamma^5 u_s(q_1)  \ .
\end{align}
where $\epsilon_{\nu}(\omega)$ is the polarization of produced $\omega$ meson and
\begin{align}
F_{f_1NN}(t)= \frac{1}{(1 - t/M^2_{f_1})^2} \ .
\end{align}
In the unpolarized case, the differential cross section for the $\omega$-photoproduction is:
\begin{align}
\frac{d\sigma^{\omega}_{f_1}}{dt} &= \frac{A_f F^2_{f_1NN}(t)}{8(t - M^2_{f_1})^2} \bar{\Phi}_s(s,t) \ ,
\end{align}
where $ A_{f} \equiv \alpha_{\rm em} \ g^2_{f_1NN} \ g^2_{\gamma \omega f_1} $ and
\begin{align}
\bar{\Phi}_s(s,t) = \frac{M^2_N- u}{s - M^2_N}
+ \frac{(M^2_{\omega} - t)}{2M^2_{\omega}}\left[\frac{(M^2_N - u)(M^2_N + M^2_{\omega} - u)}{(s - M^2_N)^2} + \frac{(s - M^2_N - M^2_{\omega})}{(s - M^2_N)} + 2M^2_N\frac{(M^2_{\omega} - t)}{(s - M^2_N)}\right] \ .
\end{align}
Similarly, for $\rho$-photoproduction, we will have
\begin{align}
\frac{d\sigma^{\rho}_{f_1}}{dt} &= 9\frac{d\sigma^{\omega}_{f_1}}{dt}  \ ,
\end{align}
where we took into account that $M_{\omega} = M_{\rho}$ and $ g_{\gamma \rho f_1} = 3g_{\gamma \omega f_1} $.


To Reggeize the propagator, we make the following substitution:
\begin{align}\label{Reggeprop2}
\frac{1}{t - M^2_{f_1}} \ \rightarrow \   \frac{\pi \ \alpha'_{f_1}}{2 \ \Gamma(\alpha_{f_1}(t))} \ \eta(t)\left(\frac{s}{s_0}\right)^{\alpha_{f_1}(t)-1}  \  ,
\end{align}
therefore, for large $s$,
\begin{align}
\frac{d\sigma^{\omega}_{R}}{dt} &= \pi^2 A_f F^2_{f_1NN}(t) \ \frac{\left(1 - \frac{t}{2M^2_{\omega}}\right)\alpha^{'2}_{f_1}}{16 \ \Gamma^2( \alpha_{f_1}(t)) \cos^2\left(\frac{\pi \alpha_{f_1}(t)}{2}\right)} \ \left(\frac{s}{s_0}\right)^{2(\alpha_{f_1}(t)-1)} \ .
\end{align}
For $\rho$ photoproduction, the differential cross section  will be $9$ times larger.
Exchange of  the $f_1$ Regge trajectory  in $\omega$ or $\rho$ photoproduction leads to the same
polarization structure as found previously for $f_1$ photoproduction via the exchange of the $\omega$ and
$\rho$ Regge trajectories. This structure may be useful in trying to unravel the unknown Regge trajectory for the $f_1$ meson from measurements of polarized $\rho$ and $\omega$ photoproduction.


\section{Conclusions}

Working in the theoretical framework of holographic QCD, we have calculated the couplings between vector and axial-vector mesons, the pion,
and the photon which emerge from the 5D Chern-Simons term.
Though interactions with the same structure were studied before (see, e.g. Ref.~\cite{Kochelev:1999zf}),
the advantage of holographic QCD is that it allows us to derive these interactions from a unified framework and determine all
 possible couplings between the mesons and the photon.
Both holographic models we employ have only two free parameters, the gauge coupling strength ($g_5$ or $\lambda$) and the confinement scale ($1/z_0$ and $M_{KK}$), which are fixed by the  $\rho$ meson mass and the pion decay constant.

To examine the experimental manifestation of these interactions, we studied the photoproduction of vector and axial
vector mesons, such as $\omega$, $\rho$ and $f_1$, with special emphasis on $f_1$ photoproduction, as this is expected to
have the clearest experimental signature.
After determining the vertex functions $K_{\gamma V f_1}(t)$ and couplings
from holographic QCD, we borrowed the form factors and couplings involving the nucleon from the
phenomenologically accepted models (including other holographic models) to calculate the scattering amplitudes at low energies (where
single-particle exchange is a good approximation), then at large $s$ and small $|t|$, where we used Regge theory to substitute
the single meson propagator with its Reggeized version.
This yielded predictions for the differential cross sections for $f_1$ photoproduction in the $t$-channel. After considering other
processes as well, we show that both the contribution from nucleon exchanges in the $u$- and $s$-channels and double photoproduction
can be neglected.
The primary contribution to the $f_1$ photoproduction comes from the exchange with $\omega$ and $\rho$ mesons. For large energies, the Regge trajectories of these mesons begin to dominate the process.

We consider both unpolarized and polarized cases and show that in the polarized case the polarization of the produced meson must be perpendicular to the polarization of the incident photon. This is specific to the ``anomalous'' type of interactions that contain a Levi-Civit\'{a} symbol. Precisely for this reason, polarized $\omega$ or $\rho$ photoproduction may be a very useful tool to determine the slope and intercept of the $f_1$ meson Regge trajectory.

Finally, we mention some interesting directions for future work. Our focus has been on hadronic couplings to the photon,  it would be natural to generalize our work to include the electroweak gauge bosons $W^\pm, Z^0$ as background fields and to compare to the results of  \cite{Harvey:2007ca}. In our treatment of photoproduction we have taken the nucleon couplings and form factors as determined from other phenomenological fits. It would be more consistent to calculate these in a holographic framework  using previous results on the description of nucleons in string duals of QCD
\cite{Hata:2007mb, Hashimoto:2008zw, Hong:2007ay}.  Our calculations have been done  for $N_f=2$ and in the chiral limit of massless pions. The extensions to $N_f=3$ and to massive quarks are worth pursuing. Finally, it would be interesting to see if there are other photoproduction processes that can be used to further probe the types of couplings that we have analyzed here.

{\bf Note Added:} As this work was being finished a preprint appeared which also considers photoproduction of $f_1$ mesons at JLab \cite{Kochelev:2009xz}. Their predicted differential cross section differs from ours.

\begin{acknowledgements}
JH would like to thank J. Dudek, C. Hill, R. Hill and J. Rosner for a number of helpful discussions. HG thanks T.~S.~Lee for valuable comments and the EFI for hospitality. This work was supported in part by
NSF Grants PHY-00506630 and 0529954 and by DOE ONP contract DE-AC02-06CH11357.

\end{acknowledgements}



\appendix

\section{Three- and four-point pseudo-Chern-Simons Couplings}

\subsection{The Hirn-Sanz model}

We summarize the couplings resulting from the Chern-Simons action in the Sakai-Sugimoto and Hirn-Sanz models.  A number of these couplings have been worked for $N_f=3$ in the Sakai-Sugimoto model  \cite{Sakai:2005yt}.

Recall the Chern-Simons action for the Hirn-Sanz model
\begin{equation}
S_{CS}=\frac{N_c}{24\pi^2} \int \left[\omega_5(L)-\omega_5(R)\right] \ ,
\end{equation}
with
\begin{eqnarray}
\omega_5(A)&=& \tr\left[ AF^2+\frac{i}{2}A^3F-\frac{1}{10}A^5\right] \\
&=& \tr\left[ A(dA)^2-\frac{3i}{2}A^3dA-\frac{3}{5}A^5\right] \ ,
\end{eqnarray}
where $L$ and $R$ are $U(N_f)$-valued. The Chern-Simons (CS) action is only defined up to boundary terms. As in
\cite{Sakai:2004cn}, we assume that the action is initially written in a gauge where all gauge fields vanish
at the (UV) boundary of the space. When we transform to $L_z=R_z=0$ gauge, we generate boundary terms
of the form
\begin{equation}
S_{{\rm bdy}} = \frac{N_c}{24\pi^2}\int_{M^4}\left[\alpha_4(d\xi_L^{-1}\xi_L, L)-\alpha_4(d\xi_R^{-1}\xi_R, R) \right] \ ,
\end{equation}
where $\alpha_4$ is
\begin{equation}
\alpha_4(V,A)=-\frac{i}{2}\tr\left(V(iAdA + idAA + A^3) - \frac{1}{2}VAVA - V^3A\right)~.
\end{equation}
(There is also a WZW-like term which yields 5- and higher-point couplings.)
Writing $V=(L+R)/2$ and $A=(L-R)/2$ as usual, we note that only terms containing an odd
number of $A$'s will survive in the action, giving
\begin{equation}
S_{CS}=S_{CS-3}+S_{CS-4}+S_{CS-5} \ ,
\end{equation}
where
\begin{eqnarray}\label{eq:CSterms}
S_{CS-3}&=&\frac{N_c}{12\pi^2}\int \tr\left[ A(dA)^2+V dA dV + V dV dA+A(dV)^2\right] \ , \\
S_{CS-4}&=&-\frac{iN_c}{8\pi^2}\int \tr\left[\left(V^2A+AV^2\right)dV+VAV dV+V^3dA\right]+\left( V\leftrightarrow A\right) \ , \\
S_{CS-5}&=&-\frac{N_c}{4\pi^2}\int\tr\left[V^4 A+V^2A^3+3VAVA^2+\frac{1}{5}A^5\right]~
\end{eqnarray}
and similarly for the boundary terms from $\alpha_4$, which contribute exclusively photon-pion couplings.
We now expand the new (gauge-transformed) fields $V_\mu$ and $A_\mu$,  keeping only the vector source $\hat{V}$, the pion $\pi$, and
the lightest (axial) vector meson states:
\begin{align}
V_\mu(x,z) &= \hat{V}_\mu(x) + \frac{i}{2f_\pi^2}[\pi,\dt_\mu\pi] + \rho_\mu(x)\psi_{\rho}(z) + \dots \ , \\
A_\mu(x,z)&=\alpha(z)\frac{1}{f_\pi}\dt_\mu\pi - i\frac{\alpha(z)}{f_\pi}[\hat{V}_\mu,\pi] + a_\mu(x)\psi_{a_1}(z) + \dots \ ,
\end{align}
where we have defined $\xi_R(x)=\xi_L^\dagger(x)=e^{i\pi/f_\pi}$ and expanded to second order in the fields.
Note that we have absorbed the normalizable pieces of the gauge transformations $\tilde{\xi}_L$ and $\tilde{\xi}_R$ into
the definitions of the vector mesons.
The story is almost identical for the Sakai-Sugimoto model, where there is a single gauge field $A_\mu$ which splits into
axial and vector pieces, as discussed above. Boundary terms are generated by the presence of two UV boundaries (L and R) which can similarly
be exchanged by parity. The terms living on the brane (not at the boundaries) thus differ by an overall factor of $1/2$ with respect
to the Hirn-Sanz model terms given in \ref{eq:CSterms}.

We work with $N_f=2$, with  generators of $U(2)$ normalized such that ${\rm Tr} T^aT^b=\delta^{ab}/2$ (for $a,b =0,1,2,3$).
%
As has been noted before, the photon is generated by the EM current (\ref{eq:EMcurrent}).
The photon appears in overlap integrals as the (nonnormalizable) zero-momentum solution to the vector gauge
field equations of motion -- in this case, $\psi_\gamma=1$.
Also, each occurrence of the photon field $\gamma_\mu$ in the tables below should be accompanied by a factor of the
electromagnetic coupling $e$. We give the couplings in terms of overlap integrals between wavefunctions using the notation for Hirn-Sanz
\begin{equation}
I_{B'CCD}=\int_0^{z_0} dz\psi_B' \psi_C^2\psi_D~,
\end{equation}
and for Sakai-Sugimoto
\begin{equation}
I_{B'CCD}=\frac{1}{2}\int_{-\infty}^{\infty} dz\psi_B' \psi_C^2\psi_D~ = \int_0^{\infty} dz\psi_B' \psi_C^2\psi_D~.
\end{equation}
Because the overall coefficient of bulk integrals in Sakai-Sugimoto differs by a factor of $1/2$ with respect to
Hirn-Sanz, but the domain of integration is $(-\infty,\infty )$, this notation allows us to write the couplings in the two models in precisely the same form. The wavefunctions are normalized to yield the canonical kinetic terms in the Yang-Mills Lagrangian.

To use these tables to extract the listed couplings one should take the interaction term
given in the second column, contract the Lorentz indices with  $\epsilon^{\mu\nu\rho\sigma}$ and multiply by $N_c/48 \pi^2$, then write the resulting interaction term as the coupling
given in the first column times terms involving fields and derivatives and the Levi-Civita tensor, but no constants.  For example, the
$\rho-f_1-\gamma$ interaction is $-6 I_{AV'}(N_c/48 \pi^2) \epsilon^{\mu \nu \rho \sigma} f_\mu \partial_\nu \gamma_\rho \rho^0_\sigma$ which we rewrite as $g_{\gamma \rho f}  \epsilon^{\mu \nu \rho \sigma} f_\mu \partial_\nu \gamma_\rho \rho^0_\sigma$ with $g_{\gamma \rho f}= -6 I_{AV'}(N_c/48 \pi^2)$. The numerical values of these couplings are given in the final two columns of the tables, computed in the HS and SS models, respectively. We fix the free parameters in both the Hirn-Sanz (HS) and Sakai-Sugimoto (SS) models using $m_\rho$ and $f_\pi$.


\begin{center}
\begin{tabular}{|c|c|c|c|}
\multicolumn{4}{c}{TABLE III: Three-point couplings.}\\  \hline
 g &   Interaction $ \times (N_c/48\pi^2)^{-1}$ & Value in HS & Value in SS \\ \hline\hline
$g_{\gamma\rho f}$ & $ -6I_{AV'} f_\mu\dt_\nu\gamma_\rho\rho^0_\sigma$& $-1.03$ & $-0.95$\\ \hline
$g_{\gam\gam\pi}$ & $2 f_\pi^{-1}\gamma_\mu\dt_\nu\gamma_\rho\dt_\s\pi^0 $ & $-0.01f_\pi^{-1}$ & $-0.01f_\pi^{-1}$\\ \hline
$g_{\gam\rho a}$ & $-2I_{AV'}a_\mu^a\dt_\nu\gamma_\rho\rho_\s^b\delta_{ab}$ & $-0.34$  & $-0.32$  \\ \hline
$g_{\gam\rho\pi }$ & $2I_{\pi'V}f_\pi^{-1}\dt_\mu\pi^a\dt_\nu\gamma_\rho\rho_\s^b\delta_{ab}$ & $-0.06f_\pi^{-1}$  & $-0.06f_\pi^{-1}$ \\ \hline
$g_{\gam a\om}$ & $-6I_{AV'}a_\mu^0\dt_\nu\gamma_\rho\om_\s$ & $-1.03$  & $-0.95$ \\ \hline
$g_{\gam \pi\om}$ & $6I_{\pi' V}f_\pi^{-1}\dt_\mu\pi^0\dt_\nu\gamma_\rho\om_\s$ & $-0.18f_\pi^{-1}$  & $-0.18f_\pi^{-1}$  \\ \hline
$g_{\pi a f}$ & $3I_{AA\pi'}f_\pi^{-1}\dt_\mu\pi^a(\dt_\nu a_\rho^bf_\s-\dt_\nu f_\s a_\rho^b)\delta_{ab}$ & $-0.31f_\pi^{-1}$ & $-0.26f_\pi^{-1}$ \\ \hline
$g_{\gam\om f}$ & $-2I_{AV'}f_\mu\dt_\nu\gam_\rho\om_\s$ & $-0.34$  & $-0.32$ \\ \hline
$g_{\om \rho a}$ & $-3I_{A'VV}a_\mu^a(\dt_\nu\om_\rho\rho_\s^b-\om_\rho\dt_\nu\rho_\s^b)\delta_{ab}$ & $4.09$  & $3.70$  \\ \hline
$g_{\om\rho\pi }$ & $-3I_{\pi'VV}f_\pi^{-1}\dt_\mu\pi^a(\dt_\nu\om_\rho\rho_\s^b-\om_\rho\dt_\nu\rho_\s^b)\delta_{ab}$ & $-0.52f_\pi^{-1}$  & $-0.51f_\pi^{-1}$  \\ \hline
$g_{\rho\rho f}$ & $3I_{A'VV}f_\mu\dt_\nu\rho_\rho^a\rho_\s^b\delta_{ab}$ & $-4.09$ & $-3.70$\\ \hline
$g_{\om\om f}$ & $3I_{A'VV}f_\mu\dt_\nu\om_\rho\om_\s$ & $-4.09$  & $-3.70$ \\ \hline
\end{tabular}
\end{center}

The four-point couplings are
\begin{center}
\begin{tabular}{|c|c|c|c|}
\multicolumn{4}{c}{TABLE IV: Four-point couplings.}\\  \hline
   g &   Interaction $\times (N_c/48\pi^2)^{-1}$& Value in HS & Value in SS \\ \hline\hline
 $g_{\gam\gam\rho\pi}$ &  $-2I_{\pi V'}f_\pi^{-1} \gam_\mu\dt_\nu\gam_\rho\pi^a\rho_\rho^b\eps^{ab3}$ & $-0.06f_\pi^{-1}$ & $-0.06f_\pi^{-1}$ \\ \hline
$g_{\gam\rho\rho f}$ & $3I_{A'VV}\rho_\mu^a\rho_\nu^bf_\rho\gam_\s\eps^{ab3}$ & $4.09$ & $3.70$ \\ \hline
$g_{\gam\rho\om a}$ & $3I_{A'VV}\gam_\mu\rho_\nu^aa_\rho^b\om_\s\eps^{ab3}$ & $4.09$ & $3.70$\\ \hline
$g_{\gam\rho\om \pi}$ & $-3I_{\pi'VV}f_\pi^{-1}\gam_\mu(\dt_\nu\om_\rho\pi^a\rho_\s^b-\pi^a\om_\rho\dt_\nu\rho_\s^b+\dt_\nu\pi^a\om_\rho\rho^b_\s)\eps^{ab3}$ & $0.52f_\pi^{-1}$ & $0.51f_\pi^{-1}$\\ \hline
$g_{\gam \pi \pi f}$ & $3I_{A'\pi\pi}f_\pi^{-2}\eps^{ab3}\dt_\mu\pi^a\dt_\nu\pi^b f_\rho \gam_\s$ & $-0.08f_\pi^{-2}$ & $-0.07f_\pi^{-2}$\\ \hline
$g_{\gam \pi af}$ & $3I_{AA\pi'}f_\pi^{-1}\gam_\mu(\pi^a\dt_\nu a_\rho^bf_\s-\pi^a\dt_\nu f_\s a_\rho^b+\dt_\nu\pi^aa_\rho^bf_\s)\eps^{ab3}$ & $-0.31f_\pi^{-1}$ & $-0.26f_\pi^{-1}$\\ \hline
$g_{\gam \pi\pi\pi}$ & $ (f_\pi^{-3}/2)\dt_\mu\pi^a\dt_\nu\pi^b\dt_\rho\pi^c\gam_\s\eps^{abc} $ & $0.003f_\pi^{-3}$ & $0.003f_\pi^{-3}$\\ \hline
$g_{\rho\rho\rho f}$ & $I_{A'VVV}f_\mu\rho_\nu^a\rho_\rho^b\rho_\s^c\eps^{abc}$ & $9.52$ & $8.42$ \\ \hline
$g_{\om\rho\rho a}$ & $I_{A'VVV}\rho_\mu^a\rho_\nu^b a_\rho^c\om_\s\eps^{abc}$ & $9.52$ & $8.42$ \\ \hline
$g_{\om\rho\rho \pi}$ & $I_{\pi' VVV}f_\pi^{-1}\rho_\mu^a\rho_\nu^b\dt_\rho\pi^c\om_\s\eps^{abc}$ & $-1.03f_\pi^{-1}$ & $-1.03f_\pi^{-1}$ \\ \hline
$g_{\om aaa}$ & $I_{AAA V'}a_\mu^a a_\nu^b a_\rho^c\om_\s \eps^{abc}$ & $3.78$ & $2.93$\\ \hline
$g_{\om \pi aa}$ & $3I_{AA\pi V'}f_\pi^{-1}a_\mu^a a_\nu^b\dt_\rho\pi^c\om_\s\eps^{abc}$ & $1.49f_\pi^{-1}$ & $1.31f_\pi^{-1}$\\ \hline
$g_{\om \pi \pi a}$ & $3(I_{AV'}+I_{A\pi\pi V'})f_\pi^{-2}a_\mu^a\dt_\nu\pi^b\dt_\rho\pi^c\om_\s\eps^{abc}$ & $0.75f_\pi^{-2}$ & $0.70f_\pi^{-2}$\\ \hline
$g_{\om \pi \pi \pi}$ & $(3I_{\pi V'}-I_{V'\pi\pi\pi})f_\pi^{-3}\dt_\mu\pi^a \dt_\nu\pi^b\dt_\rho\om_s\pi^c\eps^{abc}$ & $-0.07f_\pi^{-3}$ & $-0.07f_\pi^{-3}$\\ \hline
$g_{\rho aaf}$ & $I_{AAAV'}f_\mu a_\nu^aa_\rho^b\rho_\s^c\eps^{abc}$ & $3.79$ & $2.93$\\ \hline
$g_{\rho \pi af}$ & $-6I_{AA'\pi V}f_\pi^{-1}\dt_\mu\pi^a a_\nu^b\rho_\rho^c f_\s\eps^{abc}$ & $0.04f_\pi^{-1}$ & $0.13f_\pi^{-1}$\\ \hline
$g_{\rho \pi\pi f}$ & $3(I_{A'V}+I_{A'\pi\pi V})f_\pi^{-2}f_\mu\dt_\nu\pi^a\dt_\rho\pi^b\rho_\s^b \eps^{abc}$ & $-0.42f_\pi^{-2}$ & $-0.42f_\pi^{-2}$\\ \hline
\end{tabular}
\end{center}


\section{Single Nucleon Exchange Contribution to $f_1$ Photoproduction}

\begin{figure}[h]
\includegraphics[width=6.1in]{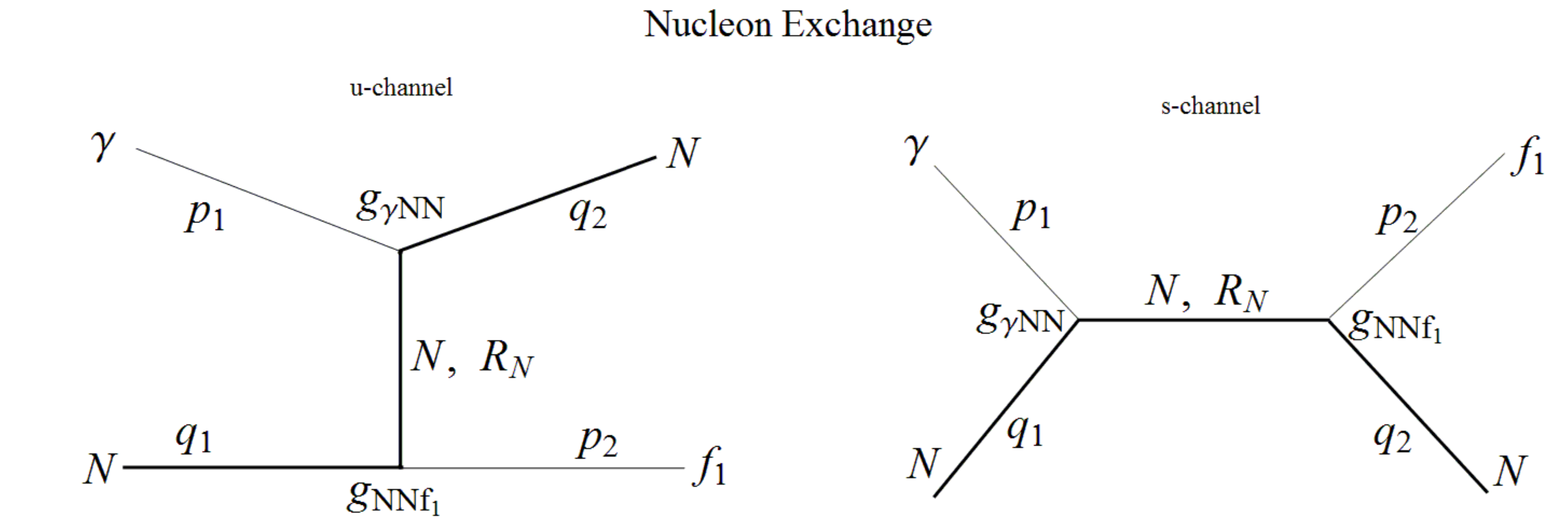}
\caption{\label{Fig3} Photoproduction of $f_1$ meson in $u$- (left) and $s$- (right) channel correspondingly, via the exchange of either single nucleon or nucleon Regge trajectory.}
\end{figure}

The interactions relevant for nucleon exchange diagrams on Fig.~\ref{Fig3} (when nucleon is proton)
emerge from the following Lagrangians:
\begin{align}
&\cL_{\gamma pp} = -e \ \bar{\psi}_N \left[\gamma^{\mu}A_{\mu} - \frac{\kappa_p}{2M_N}\sigma_{\mu\nu}\partial^{\nu}A^{\mu}\right]\psi_N \ , \\[7pt] \nonumber
&\cL_{f_1 NN}  = g_{f_1NN} \ \bar{\psi}_N \left[f_{1\mu} - i\frac{\kappa_{f_1}}{2M_N}\gamma^{\nu}\partial_{\nu} f_{1\mu}\right]\gamma^{\mu}\gamma^{5}\psi_N  \ ,
\end{align}
where $A_{\mu}$ describes electromagnetic field, $f_{1\mu}$ describes the isosinglet axial-vector meson field, $\kappa_p \simeq 1.79 $ is the anomalous magnetic moment of the nucleon. Here, we will ignore $\kappa_{f_1}$ as well as $\kappa_{p}$, since these are $q/M_N$ suppressed. The value of the coupling $g_{f_1NN} =  2.5 \pm 0.5$ is fixed through the proton spin analysis \cite{Birkel:1995ct}.

The production amplitude in $u$- and $s$- channels can be written as:
\begin{align}
\cM^{\mu\nu}_{u,s}=  e \ g_{f_1NN} \ \bar{u}(q_2) \left[
\gamma^{\mu}\frac{\sla{q_1}-\sla{p_2}+M_N}{u - M^2_N}\gamma^{\nu}\gamma^5F_N(u) +
\gamma^{\nu}\gamma^5\frac{\sla{p_1}+\sla{q_1}+M_N}{s - M^2_N}\gamma^{\mu}F_N(s)
\right]u(q_1) \ ,
\end{align}
where
\begin{align}
F_{N}(u) = \frac{\Lambda^4_N}{\Lambda^4_N + (u - M^2_N)^2} \ ,
\end{align}
and $ \Lambda_N = 0.5 \ \GeV $ taken from the Ref.~\cite{Oh:2003aw} and references therein.


When $F_N(u) = F_N(s) = F $, it follows that $ p^{\mu}_1 \cM^{\mu\nu}_{u,s} = 0 $ when $p^2_1 = 0$.
In general, $F_N(u) \neq F_N(s) $ and the gauge invariance is broken (the term proportional to $\kappa_p $ won't change the situation). The gauge invariance may be restored by redefining the interaction vertices. However, this shouldn't significantly change the final results.

Without going into details, it suffices to show that in the limit, when $ s \gg M^2_N $ and $ t $ is fixed,
\begin{align}\label{NuclExch2}
\frac{d\sigma_N}{dt} \simeq \frac{\alpha_{\rm em} g^2_{f_1NN}}{M^4_N} \ \left(\frac{\Lambda_{N}}{M_N}\right)^8 \ \frac{1}{\tilde{s}^6} \sim \frac{0.001}{\tilde{s}^6} \ ,
\end{align}
where $ \tilde{s} = s/M^2_N $. For comparison, in the same limit, the differential cross section corresponding to
single $\omega$ and $\rho$ meson exchange is:
\begin{align}
\frac{d\sigma_V}{dt} \sim \frac{1}{\pi}\left(\frac{r \ g_{\rho NN} \ eg_{\gamma \rho f_1}}{4M^2_{\rho}}\right)^2  \left(\frac{\Lambda_{VNN}}{M_N}\right)^8 \ \frac{1}{\tilde{s}^4} \sim \frac{0.1}{\tilde{s}^4} \ .
\end{align}
Summarizing, at large $s$ and fixed (small) $|t|$, the cross section corresponding to nucleon exchange decreases as $ 1/s^6$, and the meson contribution is absolutely dominant over the nucleon contribution.

\section{$f_1\rightarrow \rho^0 + \gamma$}

As a further check of the predictions of AdS/QCD for pseudo-Chern-Simons terms we use the couplings given in Table III  to compute the rate and polarization structure for the decay $f_1 \rightarrow \rho^0 + \gamma$ \footnote{This interaction was also derived in \cite{Harvey:2007ca} based on a gauging of the Standard Model gauge symmetries in the Wess-Zumino-Witten interaction. The result here is more specific as we do not assume universality of vector meson couplings. The computation of the decay rate based on this coupling was also done in \cite{Kochelev:1999zf}.}.
The interaction Lagrangian for this process is given in
\begin{align}
\cL_{\gamma \rho^0 f_1} =  e g_{\gamma \rho f} \epsilon^{\mu\nu\lambda\rho}
\partial_{\mu}A_{\nu}\rho_{\lambda}f_{1,\rho}\ .
\end{align}
For photon momentum $q$ and polarization vectors $\epsilon^{(\gamma)}, \epsilon^{(f)}$ and $\epsilon^{(\rho)}$ for the photon, $f_1$ and $\rho$ respectively the decay amplitude is
\begin{align}
i {\cal M} = i e g_{\gamma \rho f} \epsilon^{\mu \nu \alpha \beta} \bigl( i q_\mu \epsilon_\nu^{(\gamma) *} \bigr) \epsilon_\alpha^{(\rho)*} \epsilon_\beta^{(f)}
\end{align}
and leads to a decay rate
\begin{align}
\Gamma(f_1 \rightarrow \rho^0 + \gamma) = \frac{\alpha g_{\gamma \rho f}^2}{3}  \frac{E_\gamma^3}{m_\rho^2} \biggl( 1 + \frac{m_\rho^2}{m_f^2} \biggr)
\end{align}
where $E_\gamma= (m_f^2-m_\rho^2)/2m_f$ is the energy of the final state photon in the $f_1$
rest frame.  Comparison to the measured rate \cite{pdg}
\begin{align}
\Gamma_{\rm exp}(f_1 \rightarrow \rho^0 + \gamma) = 1.33~ \pm .37~ {\rm MeV}
\end{align}
requires that $|g_{\gamma \rho f}|=1.7~ \pm .4$ which should be compared with the value
of $g_{\gamma \rho f} \simeq -1 $ given in Table III of Appendix A. The agreement is not spectacular, but is much better than an old quark model calculation of this process \cite{Babcock}. We note that if we use $g_5= 2 \pi$ in the Hirn-Sanz model as determined from the
vector-vector two point function rather then the value $g_5= 4.91$ determined from fitting to $f_\pi$ we obtain
$g_{\gamma \rho f}= -1.68$.

A study of the polarization structure shows that the ratio for the decay of longitudinal to transverse $\rho$ mesons is $\Gamma_{\rm long}/\Gamma_{\rm trans} = m_f^2/m_\rho^2 \simeq 2.7$. Unfortunately the experimental situation appears to be unclear. The papers \cite{Coffman,Amelin} give conflicting results for the polarization structure. Our results are in rough agreement with the results of \cite{Amelin}.


\end{document}